\documentclass[twocolumn]{emulateapj-rtx4}

\usepackage{amssymb}
\usepackage{lscape}
\usepackage{color}
\usepackage{graphicx}
\usepackage{makecell}
\usepackage{morefloats}

\begin{document}

\title{Variability-based AGN selection using image subtraction in the SDSS and LSST era}

\author{Yumi Choi\altaffilmark{1}, Robert R. Gibson\altaffilmark{1}, Andrew C. Becker\altaffilmark{1}, \v Zeljko Ivezi\'c\altaffilmark{1}, Andrew J. Connolly\altaffilmark{1}, Chelsea L. MacLeod\altaffilmark{2}, John J. Ruan\altaffilmark{1}, Scott F. Anderson\altaffilmark{1}}
\email{ymchoi@astro.washington.edu}

\altaffiltext{1}{Department of Astronomy, University of Washington, Box 351580, Seattle, WA 98195, USA}
\altaffiltext{2}{Physics Department, U.S. Naval Academy, 572 Holloway Rd, Annapolis, MD 21402, USA}

\shorttitle{AGN optical variability}
\shortauthors{Choi et al.} 

\hbadness=10000

\begin{abstract}
With upcoming all sky surveys such as LSST poised to generate a deep digital movie of the optical sky, 
variability-based AGN selection will enable the construction of highly-complete catalogs with minimum 
contamination. In this study, we generate $g$-band difference images and construct light curves for QSO/AGN 
candidates listed in SDSS Stripe 82 public catalogs compiled from different methods, including spectroscopy, 
optical colors, variability, and X-ray detection. Image differencing excels at identifying variable sources 
embedded in complex or blended emission regions such as Type II AGNs and other low-luminosity AGNs 
that may be omitted from traditional photometric or spectroscopic catalogs. To separate QSOs/AGNs from 
other sources using our difference image light curves, we explore several light curve statistics and 
parameterize optical variability by the characteristic damping timescale ($\tau$) and variability amplitude. 
By virtue of distinguishable variability parameters of AGNs, we are able to select them with high completeness 
of 93.4\% and efficiency (i.e., purity) of 71.3\%. Based on optical variability, we also select highly variable  
blazar candidates, whose infrared colors are consistent with known blazars. One third of them are also radio 
detected. With the X-ray selected AGN candidates, we probe the optical variability of X-ray detected 
optically-extended sources using their difference image light curves for the first time. A combination of optical 
variability and X-ray detection enables us to select various types of host-dominated AGNs. Contrary to the 
AGN unification model prediction, two Type II AGN candidates (out of 6) show detectable variability on 
long-term timescales like typical Type I AGNs. This study will provide a baseline for future optical variability 
studies of extended sources. 
\end{abstract}


\section{INTRODUCTION}
\label{sec:introSec}
Most galaxies host supermassive black holes (SMBHs) at their central regions. It is believed that
the SMBHs, especially when they are in an active or quasar mode, regulate the evolution of their 
host galaxies and their own growth through feedback processes. There is now strong evidence that 
active galactic nuclei (AGNs) shape not just their host galaxies \citep[e.g.,][]{dimatteo05, hopkins07}, 
but also the intracluster-medium in galaxy clusters \citep[e.g.,][]{begelman04}, and even the ionization 
levels of the Universe on cosmological scales \citep[e.g.,][]{loeb01}. Although the effects of AGN 
activity on larger-scales are well studied, a consensus has not been reached on the underlying physics that 
generates these effects. This is partly because we do not fully understand how AGN outflows evolve, and 
also because single-epoch studies only describe a snapshot of AGN activity. 

AGNs seem to be powered by the accretion of matter onto SMBHs with complex hydrodynamic and 
magnetic processes, along with relativistic effects \citep[e.g.,][and references therein]{springel05, 
mchardy10, choi12}. Unfortunately, we cannot directly observe the accretion disks in distant AGNs, 
as they are far too small to resolve. However, observations have shown that accretion disks surrounding 
the SMBHs often show a strong intrinsic variability in their optical/UV continuum. Optical variability seems 
to be a common trait of AGN activity and this variability can be measured out to high redshift in time-domain
imaging surveys  \citep[e.g.,][]{sesar07}. Although the physical mechanisms causing the variations in 
brightness are not yet fully understood, the intrinsic variability of AGN can provide keys to understanding its 
physical origin and the structure of their emission regions indirectly. Thus, one of the most promising approaches 
to understanding the physics and geometry of accretion processes is to examine the time variability of AGNs. 

Studies of AGN variability across the electromagnetic spectrum \citep[e.g.,][]{simonetti85, hughes92, 
trevese94, kawaguchi98, devries05, sesar07, bauer09, kelly09, kozlowski10, schmidt10, macleod11, 
butler11, kim12} have already began paving the way for future time-domain surveys such as the Large 
Synoptic Survey Telescope \citep[LSST;][]{ivezic08}, and are useful in current surveys such as 
Pan-STARRS \citep{kaiser02} and the Palomar Transient Factory \citep[PTF;][]{law09}. Optical variability 
will be an important means of identifying AGNs for upcoming imaging surveys since it is much more complete 
and efficient than the AGN selection based on optical colors alone \citep{butler11, macleod11} and far less 
expensive than spectroscopic selection. Many efforts have been made to describe the distinctive optical 
variability of AGNs (characterized by non-periodicity and long timescales of variation) and its use for identifying 
AGNs \citep[e.g.,][]{devries05}. However, until recently, it was hard to provide simple selection criteria. 
\citet{kelly09} showed that optical AGN light curves are well described by a damped random walk (first 
order continuous autoregressive) model since the empirical power spectral distribution (PSD) of AGN light 
curves shares the same form PSD($f$) $\propto$ 1/$f^{2}$, where f is frequency, with the model. 
In this model, AGN light curves can be simply parameterized by a characteristic damping timescale and 
amplitude of variability. With a larger sample of light curves, \citet{kozlowski10} tested this model and 
concluded that it is robust and efficient for AGN selection. 
A recent study by \citet{macleod11} showed that introduction of the damping timescale, $\tau$, into 
quasar selection can boost both completeness and efficiency. Furthermore, this photometric variability-based 
selection performs well even in the mid-z range (2.2 $\lesssim$ z $\lesssim$ 3.5), where optical color selection 
performs poorly \citep{schmidt10}.

AGN variability has also been studied in the X-ray, with recent studies characterizing typical variation amplitudes 
\citep[e.g.,][]{gibson12}. The connection between optical/UV and X-ray emission provides important clues to 
the emission mechanism. \citet{gibson08} suggested the physical mechanism causing the global Baldwin effect 
\citep[anti-correlation between the C IV emission line EW and the UV luminosity;][]{baldwin77} is also 
responsible for the X-ray emission. By incorporating information from the X-ray band, we will also be able to 
unveil any connections between the optical and X-ray emitting regions.

In this work, we extend the findings of earlier photometric studies on quasar variability to a difference image 
analysis of Sloan Digital Sky Survey \citep[SDSS;][]{york00} Stripe 82. Earlier studies were necessarily 
restricted to more luminous quasars since photometry is unreliable when the source cannot be modeled 
accurately due to blended galaxy emission. Our SDSS analysis opens up a new regime for difference image 
studies due to the size and coverage of Stripe 82. While there are additional challenges to using difference 
images (described in Section~\ref{sec:sDSSDiffImSec}), this method has the distinct advantage that it can be 
applied equally well to quasars and also AGNs surrounded by host-galaxy emission. Except for a few individual 
observations of nearby AGNs, previous studies generally have used ensembles of highly-luminous quasars that 
dramatically outshine their host galaxies. Of course, quasars represent only the brightest cases of AGN, a 
relatively small fraction of AGN overall. Many more AGNs have lower luminosities and are surrounded by 
significant galaxy emission. 
In order to truly study AGN demographics, influence, and evolution, we must expand our samples to this 
regime. Difference images can also be generated quickly in order to detect rapid brightness changes in near 
real-time. For these reasons, image differencing pipelines are integral to the data processing systems for 
imaging surveys such as the LSST and Pan-STARRS even if image differencing will be most useful for 
low redshift (z $<$ 0.5) objects only.

This paper is laid out as follows. In Section~\ref{sec:CatDescSec}, we briefly summarize properties 
of 4 public catalogs of AGN candidates used in this study. Section~\ref{sec:obsRedSec} gives a description 
of observations and our data reduction of SDSS. In Section~\ref{sec:Analysis}, we describe the properties 
of the AGN candidate light curves in terms of the optical variability. We verify our methodology for 
characterizing difference image LCs, and then present the completeness and efficiency of our AGN selection 
for unresolved sources. We also examine relationships between their optical variability and X-ray properties. 
In Section~\ref{sec:resolvedSrcSec}, we discuss the results for optically--resolved, but X-ray--compact 
sources. Finally, we summarize and conclude in Section~\ref{sec:concSec}. Through out the paper, we adopt 
the cosmology H$_{0} =$ 70~km~s$^{-1}$Mpc$^{-1}$, $\Omega_{m} =$ 0.3, and 
$\Omega_{\Lambda} =$ 0.7.

\section{CATALOGS OF CANDIDATE AGNs}
\label{sec:CatDescSec}
\begin{figure} [tbp]
 \begin{center}
      \includegraphics[trim=-3mm 0mm 5mm 10mm, clip, height=9cm]{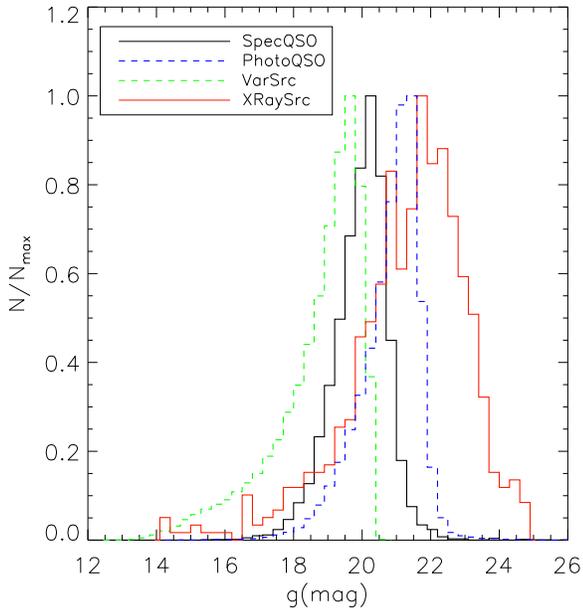}
      \caption{\label{fig1} Normalized distribution of $g$-band PSF magnitude for each catalog 
      (SpecQSO, PhotoQSO, VarSrc, XRaySrc). 
      Total numbers of sources in each photometric catalog are listed in Table~\ref{catTable}. 
      The mean magnitudes are 19.93, 20.80, 18.72, and 21.33, for SpecQSO, PhotoQSO, VarSrc, 
      and XRaySrc, respectively. From the XRaySrc catalog, here we present only point-like sources in 
      the optical.}
   \end{center}
\end{figure}

\begin{figure} [tbp]
 \begin{center}
      \includegraphics[height=10cm]{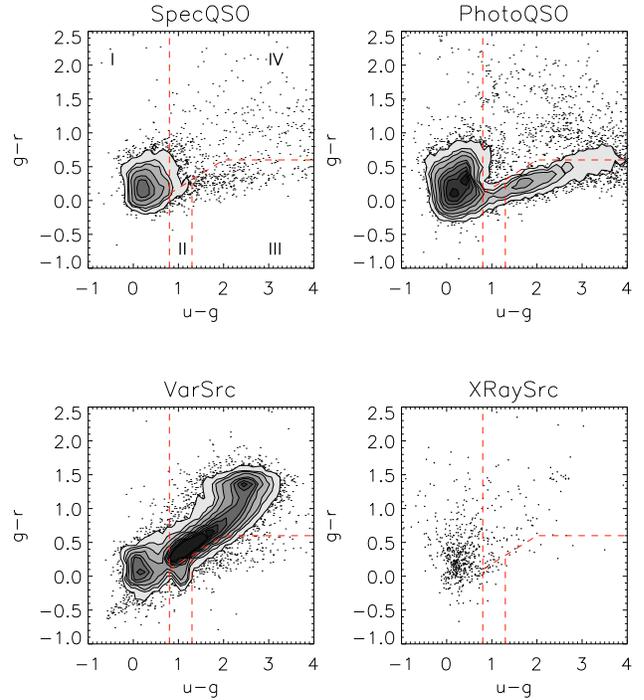}
      \caption{\label{fig2} The $u-g$ vs. $g-r$ color-color diagram of each catalog. 
      The contours show the number of data points at  
      contour levels of 10, 30, 50, 100, 200, 300, 500, 700, and 900. 
      We do not show the contour lines for XRaySrc sample.
      The characteristic regions, which are defined in the color space \citep{sesar07}, are divided by 
      the red dashed lines in each panel.}
   \end{center}
\end{figure}

Many efforts have been made to compile large samples of quasars (extremely luminous AGNs) using a 
variety of approaches, including spectroscopy \citep[e.g.,][]{schneider10}, multi-band colors \citep[e.g.,][]{richards09}, optical variability \citep[e.g.,][]{sesar07, schmidt10, macleod11, butler11}, and X-ray detections 
\citep[e.g.,][]{brandt05, cardamone08}. Therefore, we utilize 4 existing catalogs of AGN candidates 
(for details, see Table~\ref{catTable} and the Appendix) compiled by each method, each with different 
degrees of purity, to determine what AGNs look like in a difference image survey. First, we use a catalog of 
spectroscopically identified quasars (hereafter ``{\bf SpecQSO}''), which serves as a test-bed for comparison 
between difference image and photometric LC analysis, and for examining the capability of our difference image 
techniques to identify AGNs. Second, we use a catalog of quasar candidates selected by optical colors 
(hereafter ``{\bf PhotoQSO}'') to extend our tests to fainter sources and verify explicitly how well difference 
image techniques work in crowded fields, where multi-object spectroscopy suffers from fiber collision. 
Next, we use the Stripe 82 variable source catalog (hereafter ``{\bf VarSrc}'') that contains many different 
types of variable objects, including quasars, to provide a sense of how well AGNs can be separated from other 
variables in terms of temporal variability. Finally, we use a catalog of X-ray detected AGN candidates 
(hereafter ``{\bf XRaySrc}'') to examine X-ray properties with optical variability. This enables us to explore, 
for the first time, the variability of optically-extended (but compact in the X-ray) sources using difference image 
techniques. 
Before doing any analysis, we trim these catalogs to only select objects confined to the SDSS Stripe 82 
(Section~\ref{sec:obsRedSec}). Table~\ref{dupTable} shows the fractions of sources which overlap 
between trimmed catalogs. To provide a sense of the physical characteristics of each catalog, we show 
the distribution of $g$ band magnitude in Figure~\ref{fig1} and the $g-r$ vs. $u-g$ color-color diagram 
(CCD) in Figure~\ref{fig2}. Photometric information used in this study is from our Stripe 82 
dataset (Section~\ref{sec:sDSSPhotoSec}). The color-color space is divided into four characteristic 
regions (red dashed lines in Figure~\ref{fig2}): Region I for low-redshift quasars, Region II for RR 
Lyrae stars, Region III for high-redshift quasars, and Region IV for stellar locus stars \citep{sesar07}. 
In the Appendix, we describe the origins and limitations of each catalog in more detail. 

% -------------- Fig 1 -----------------

%-------------- Fig 2 -----------------

\section{OBSERVATIONS AND DATA REDUCTION}
\label{sec:obsRedSec}
The dataset for our difference image analysis comes from SDSS.
The SDSS Stripe 82 imaging data cover a large area of sky, 275 deg$^{2}$ (-50$\arcdeg < 
\alpha_{J2000.0} < 60\arcdeg$ and $|\delta_{J2000.0}| < 1.25\arcdeg$), with repeated observations 
over a decade in five bandpassess \citep[$ugriz$;][]{fukugita96}. Its repeated observations and accurate 
photometry have enabled variability studies of RR Lyrae, supernovae, quasars, blazars, etc. 
\citep[e.g.,][]{sesar07, frieman08, sako08, macleod10}. 
With $\sim$70 observations per filter for $\sim$10 years, supplemented by high-quality SDSS 
spectroscopy, Stripe 82 is a good test bed for a difference image study of AGN activity. 
In this study, we use both photometric (see Section~\ref{sec:sDSSPhotoSec}) and difference 
image (see Section~\ref{sec:sDSSDiffImSec}) data from Stripe 82 $g$-band observations to 
explore AGN optical variability. We probe the $g$-band variability because shorter wavelengths generally 
show more AGN activity \citep[e.g.,][]{vandenberk04}, while the SDSS $g$-band allows deeper 
images than the $u$ band. We will characterize photometric and difference image variability for the 
sources listed in 4 different public catalogs of AGN candidates. 

\subsection{SDSS Photometry}
\label{sec:sDSSPhotoSec} 
\begin{figure} [tbp]
  \begin{center}
      \includegraphics[trim=0mm 10mm 0mm 15mm, clip, width=8.5cm]{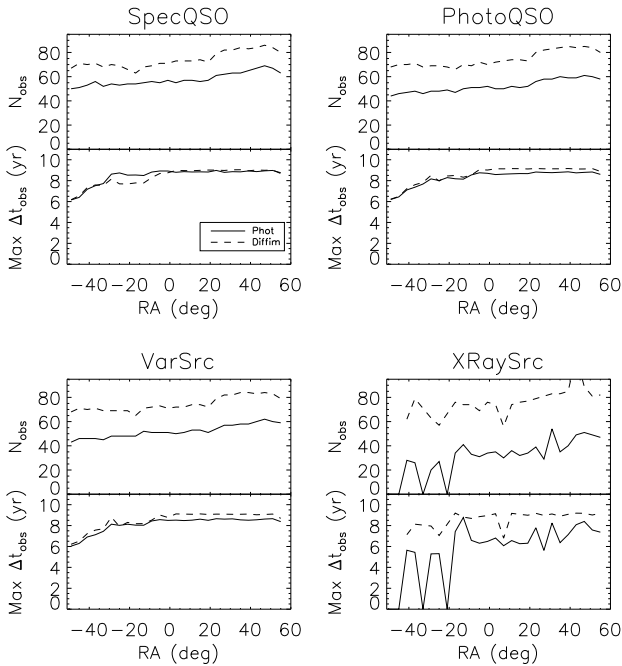}
      \caption{\label{fig3} Number of detections (top in each panel) and the maximum 
      $\Delta$\,t$_{obs}$ (bottom in each panel) as a function of right ascension (RA). 
      The average number of detections for SpecQSO, PhotoQSO, and VarSrc samples is about 
      70 epochs in difference image data, and is about 50 epochs in photometric data. However, there is no 
      significant difference in their maximum $\Delta$\,t$_{obs}$ between difference image and 	
      photometric data, except for the XRaySrc. We note that many XRaySrc were not detected at 
      many epochs or even any epochs in the optical photometric data. This is mainly because they are 
      either faint or extended in the optical bands.}
   \end{center}
\end{figure}

\citet{ivezic07} and Sesar et al. (2007, 2010) recalibrated the first and second phase of the SDSS 
(SDSS-I \& SDSS-II) multi-epoch imaging runs to improve photometric accuracy of data obtained in 
non-photometric conditions. These recalibrated data show typical photometric zero-point errors less than 
0.02 mag. In Figure~\ref{fig3}, we show the mean number of photometric measurements and the mean 
maximum observation time interval as a function of RA for each catalogs. 
On average, this photometric data set reports PSF magnitudes for $\sim$50 measurements per 
unresolved object. For each source in the catalogs (Section~\ref{sec:CatDescSec}), we construct $g$-band 
light curves (LCs) by matching all detections of each source across all runs in the recalibrated dataset with a 
1$\arcsec$ matching radius. These single-epoch catalog based LCs are referred 
to as {\it photometric} LCs to distinguish them from LCs constructed from difference images. For each 
photometric LC, we compute quantities such as the bias corrected sample standard deviation 
$\Sigma$, which is equivalent to the rms scatter in \citet{sesar07}, reduced $\chi^2$ 
($\chi^2_{\nu}$), and the structure function (SF) if the object is detected more than 4 times in the $g$-band 
(Section~\ref{sec:lCMetricsSec}). Over 90\% of all combined AGN candidates (SpecQSO, PhotoQSO, 
VarSrc, and XRaySrc) have at least 30 measurements in their photometric LCs.  

% -------------- Fig 3 -----------------

\subsection{SDSS Difference Image}
\label{sec:sDSSDiffImSec}
Difference images are created by subtracting a deeper, coadded ``template'' image from a new 
observation (the ``science image''). The template image is convolved with a spatially-varying 
kernel that matches it to the observing conditions of the science image. 
We refer the reader to details about the fundamentals of the image subtraction technique described
in \citet{alard98} and \citet{alard00}. Transient events and variable sources stand out in the 
difference images even when blended with extended sources such as host galaxies. This technique 
has been previously used to study quasars, supernovae, gamma-ray bursts, variable stars, microlensing
events, and asteroids. The particular implementation we use in this study is developed for the LSST pipeline.\footnote{Available at http://dev.lsstcorp.org/cgit/LSST/DMS} First, we converted the SDSS imaging data (114,653 
images across all runs in Stripe 82) into the LSST image format for use by the LSST software. A single 
image is 1489 pixels (10$'$) long and 2048 pixels wide (13$'$). We excluded 18 runs from the analysis 
that suffer from systematic issues such as improperly modeled gains for the CCD amplifiers and astrometric 
misalignments. We made use of data only in camera columns 2-5 where the astrometric information in the 
image header was sufficient for image registration at the time when our difference images were generated 
(two outer columns have higher-order astrometric distortion). 

\begin{figure} [tbp] 
  \begin{center}
      \includegraphics[width=8cm]{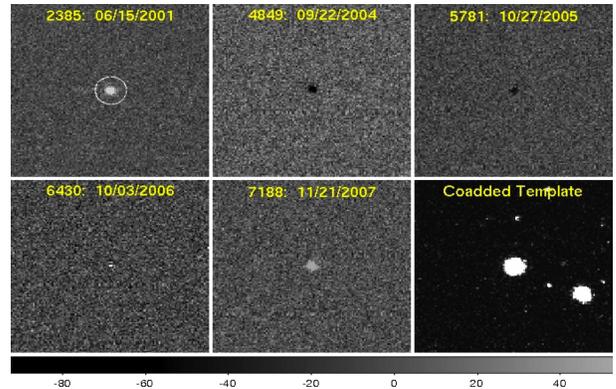}
      \caption{\label{fig4} An example of spectroscopic quasar 316.00829 +0.38036. 
      Bottom right panel is the coadded template image around the quasar and the other panels are
      difference images of it at some randomly selected epochs. Run number and the date of 
      observation are listed in each image.}
   \end{center}
\end{figure}

We use template images (artificial run number 100006 for the south strip and 200006
for the north strip) of Stripe 82 coadded by \citet{annis11}. They reach about 2 magnitudes 
deeper than a single run detection and have a median seeing of 1.1$\arcsec$ in the $r$-band.
These templates resulted from a median of the input data. Thus, the point spread function (PSF) of stars 
was seen to have a magnitude dependence, making PSF matching difficult and PSF photometry unreliable. 
For this reason, we use aperture photometry in our analysis below. 
Since any particular template image may not cover the exact same range of RA with its corresponding
science image, we stitch together two adjacent template images when we generate a difference 
image (the Dec range varies much less). Figure~\ref{fig4} shows an example of a spectroscopically confirmed 
quasar as seen in our difference images, including the coadded template image around the quasar. 
It is easy to see that residual fluxes from the object vary with time in the difference images.   

% -------------- Fig 4 -----------------

We perform forced aperture photometry in each difference image to measure residual fluxes 
for each source listed in the 4 catalogs (Section~\ref{sec:CatDescSec}). Forced aperture 
photometry is required in this case because summing over a sufficiently large aperture should 
cancel out any artificial spatial effects such as dipoles that may be produced, at the cost 
of S/N, since flux is still conserved even if the PSF matching shape is wrong. It also makes 
{\it difference image} LCs more complete since it is still possible to get a photometric data point 
from each epoch of observation even if the residual flux is negligible (but still non-zero) in a particular 
epoch difference image; correspondingly, the size of the flux measurement will be small relative to 
its error bar. By contrast, non-detections can be problematic in photometric LC analysis as 
pipelines may simply not report any epoch in which the source was not strongly detected, which
could lead to significant differences in LC sampling and, accordingly, LC metrics.

The residual fluxes for each source are summed within three apertures having radii of 2$\arcsec$, 
4$\arcsec$, and 7$\arcsec$ (the pixel scale of the SDSS is 0.396$\arcsec$). 
These apertures are seeing-corrected to reflect variation of atmospheric seeing.
No sources near the edge of an image are used in our study. If any pixel within the aperture has 
been masked, we flag the source for tracking purposes. Typically, a few masked pixels within the 
aperture do not affect the measurement of the residual flux. To make the sum of the residual fluxes 
for a given source from each difference image be comparable to one other, we divide this sum 
by the sum of the PSF matching kernels. This puts all the difference images on the photometric 
system of the common template image. The photometric uncertainty of each data point is 
estimated based on the variance plane of the science image that contains information on the total statistical 
uncertainty of a flux measurement, and then the measured uncertainty is also divided by the kernel sum. 
We empirically find that an aperture radius of 4$\arcsec$ is large enough to enclose fluxes from most sources 
in our lists, while small enough to minimize noise and contamination. Therefore, difference image LCs constructed 
with the 4$\arcsec$ aperture radius are used throughout the rest of this work. 

\begin{figure} [tbp]
 \begin{center}
      \includegraphics[trim= -10mm 0mm 0mm 20mm, clip, height=8cm]{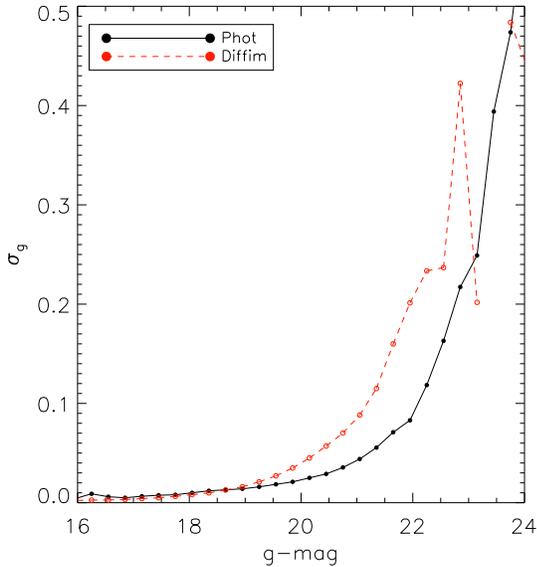}
      \caption{\label{fig5} The median uncertainties of PSF photometry (black solid) and aperture photometry 
      (red dashed) as a function of magnitude.}
   \end{center}
\end{figure}

\begin{figure} [tbp]
  \begin{center}
      \includegraphics[trim=0mm 0mm 0mm 10mm, clip, width=8cm]{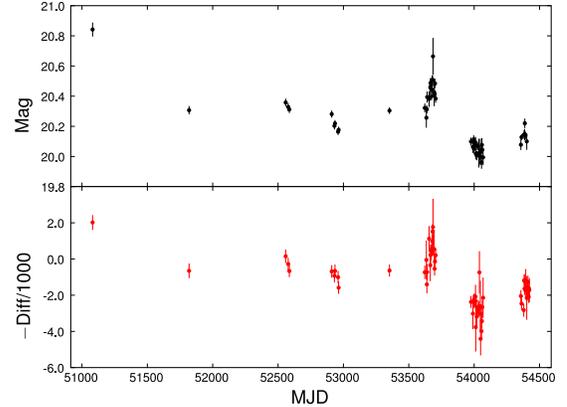}
      \caption{\label{fig6} An example comparison between a photometric LC (top) and difference
      image LC (bottom) for an object with $g \simeq$ 20.3. Although the photometric LC is in linear 
      scale and the difference image LC is in log scale, they show the same fluctuation pattern. The difference 
      image LC has relatively larger measurement errors than the photometric LC due to using aperture 
      photometry on the difference images. Note that photometric errors do not strictly 
      correlated with the source brightness since the supernova survey took data in a variety of observing 
      conditions.}
   \end{center} 
\end{figure}

In Figure~\ref{fig5}, we compare the median uncertainties of PSF photometry and our aperture 
photometry on difference images as a function of $g$ band magnitude for the SpecQSO sample. The quality of 
aperture photometry is as good as PSF photometry for bright sources ($g \lesssim$ 19) and declines with 
magnitude rapider than PSF photometry (10\% worse at $g \simeq$ 22). Since the median dispersions of our 
difference image LCs as a function of magnitude are larger than the median uncertainties by a factor of 10 (2) 
at the bright (faint) end, we argue that our difference image LCs constructed from current template images 
\citep{annis11} by performing aperture photometry are very reasonable for AGN variability study. Indeed, the 
difference image LCs reproduce their corresponding photometric LCs very well (Figure~\ref{fig6}). Furthermore, 
there are, on average, $\sim$20 more good data points in each difference image LC (dashed line in 
Figure~\ref{fig3}) than its corresponding photometric LC (solid line in Figure~\ref{fig3}). This is simply because 
the photometric recalibration was not generated for all runs in Stripe 82, while we perform forced aperture 
photometry on {\it all} difference images. However, there is no significant difference in the mean maximum 
observational time interval between the photometric and difference image data sets. This implies that, in most 
cases, missing runs in the photometric data sets are not from either the oldest or newest observations. We test 
the effect of missing data points on quantifying variability in Section~\ref{sec:diffimAdvLimSec}.   

%-------------- Fig 5 -----------------

% -------------- Fig 6 -----------------

In order to better understand the properties of difference image LCs, we also construct a {\it control}
sample of LCs generated from 10,000 randomly-chosen positions on the sky. Since the randomly 
chosen spots are essentially empty spots on the difference images, they could represent either 
non-variable sources or more likely true empty spots on the sky. With the aid of the control sample, 
we are able to provide a quantitative dividing line between non-variable and variable sources, or even 
between AGNs and other variable sources (Section~\ref{sec:identifyAGNByVarSec}).

\section{ANALYSIS OF UNRESOLVED SOURCES}
\label{sec:Analysis}
In this section, we describe our metrics and methodology to characterize LCs of all AGN candidates 
and present results for optically-unresolved sources.

\subsection{Light Curve Metrics}
\label{sec:lCMetricsSec}
To characterize AGN variability both in photometric and difference image LCs, we employ several 
LC statistics such as the $\Sigma$, $\chi^{2}_{\nu}$, and the first-order SF defined as
\begin{eqnarray}
\Sigma &=& \sqrt{\frac{1}{N-1}\,\sum_{i=1}^{N}\,(x_i - \overline x)^2}\,, \\
\chi^2_\nu &=& \frac{1}{N-1}\,\sum_{i=1}^N\,\frac{(x_i - \overline x)^2}{\xi^2_i}\,, \\
SF(\Delta\,t) &=& \sqrt{<(x_t - x_{t+\Delta\,t})^2>}\label{obsSF}\,,
\end{eqnarray}
respectively. Here, N is the number of detections, $x_{i}$ ($x_{j}$) is magnitude or flux at time 
$t_{i}$ ($t_{j}$), $\xi_{i}$ is the error on the measurement, $\overline x$ is the error-weighted 
mean measurement value, 
$\overline x = \frac{\sum_{i=1}^N\,x_i/\xi^2_i}{\sum_{i=1}^N\,1/\xi^2_i}$, and 
$\Delta\,t$ is the time lag in the {\it observer's} frame. Using the time lag in the quasar 
rest frame would require either spectroscopic observations or some estimation of photometric redshift. 
If object redshift is not known and one is only interested in selecting quasars from other sources 
based on their optical variability, the use of the time lag in the observer's frame is sufficient \citep{schmidt10, 
macleod12}. The SF analysis has been widely used to quantify AGN variability and constrain its 
variability models from observations with minimal assumptions, even when the data contain irregular 
temporal gaps \citep[e.g.][]{simonetti85, hughes92, kawaguchi98, devries05}. 

We first examine the distributions of $\Sigma$ and $\chi^{2}_{\nu}$ measured from both photometric and 
difference image LCs, and then present the SF analysis in the following section.

%-------------- Fig 7 -----------------
\begin{figure} [tbp]
 \begin{center}
      \includegraphics[trim=-10mm 0mm 0mm 10mm, clip, height=10cm]{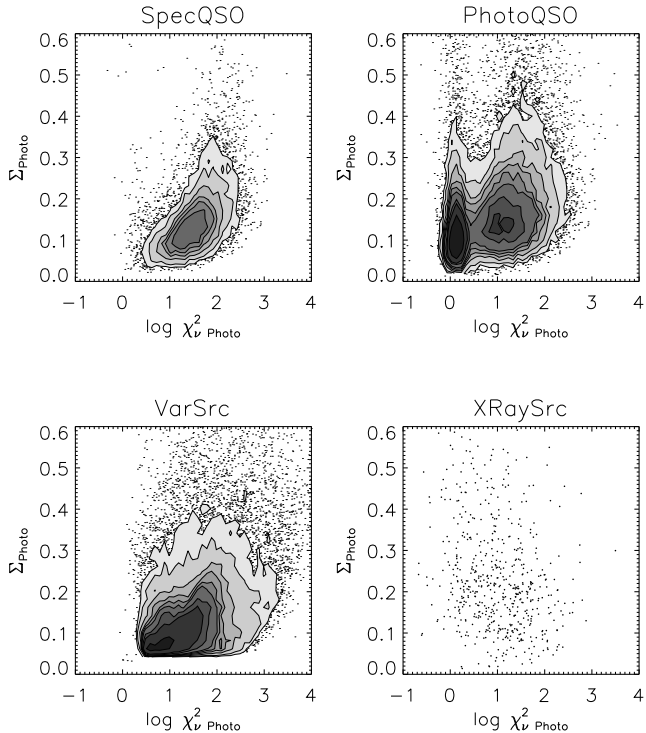}
      \caption{\label{fig7} $\Sigma$ vs. $\chi^{2}_{\nu}$ from optical photometric LCs. 
      The contours show the number of data points at contour levels of 10, 20, 40, 60, 80, 100, 150, 200, 
      and 400.}
   \end{center}
\end{figure}

In Figure~\ref{fig7}, we show the $\Sigma$ and $\chi^{2}_{\nu}$ values computed from 
photometric LCs. Since the SDSS photometric error distribution is well described by a Gaussian 
distribution \citep{ivezic07}, large $\chi^{2}_{\nu}$ values can be used to select intrinsically variable sources. 
By placing selection cuts on the $\chi^{2}_{\nu}$ and the rms scatter ($\Sigma$ in this work), 
\citet{sesar07} were able to reduce the fraction of false positives at the faint end where non-variable 
sources with large photometric uncertainties could mimic intrinsically variable sources with large values of rms 
scatter. Their selection cuts for variable sources were $\Sigma \ge$ 0.05 and  $\chi^{2}_{\nu} \ge$ 3 
in both the $g$- and $r$-bands. About 97.9\% of the SpecQSO sample (top left panel) pass the $g$-band 
selection cuts, confirming that a large fraction of quasars exhibit intrinsic variability in the optical. 
We note that SpecQSO objects with larger $\chi^{2}_{\nu}$ generally tend to have larger $\Sigma$, 
and only a few objects have log ($\chi^{2}_{\nu}$) $\gtrsim$ 2.7. This sharp edge is set by the bright 
sources in the catalog.

On the other hand, there are many objects in the PhotoQSO sample showing $\chi^{2}_{\nu} <$ 3 
(about 41\%) and there is a clear bimodality in the log ($\chi^{2}_{\nu}$) distribution around 
$\chi^{2}_{\nu} =$ 3. This is not surprising, as these objects are either possible
contaminants by non-variables or faint variables, and so their observed $\Sigma$ are likely dominated by 
Gaussian photometric noise. Since larger photometric scatter for faint non-variables can produce large 
$\Sigma$ values, we conclude that $\Sigma$, which is not normalized by uncertainty, by itself 
can not provide a good criterion for selecting intrinsically variable objects. The same trend between 
$\Sigma$ and $\chi^{2}_{\nu}$, seen in the SpecQSO sample, should occur for brighter sources only 
($g$ less than the mean magnitude of the PhotoQSO catalog), since fainter sources in the PhotoQSO 
catalog make the $\Sigma$ distribution broader.  
 
Like the SpecQSO sample, most VarSrc objects ($\sim$96.4\%) have $\Sigma \ge$ 0.05 and 
$\chi^{2}_{\nu} \ge$ 3. A few VarSrc objects, however, do not pass the current selection cuts while 
they all passed the cuts in \citet{sesar07}. This small discrepancy is merely due to a difference in 
the definition of the mean; we use an error-weighted mean while they used an unweighted mean. 
Although most objects in this sample show significant level of variability, the correlation between 
their $\Sigma$ and $\chi^{2}_{\nu}$ is apparently weaker than the SpecQSO sample, but still present. 
This may be because the VarSrc catalog includes several different classes of variable sources besides 
quasars. Existence of very bright sources, mostly stars, in the catalog allows large $\chi^{2}_{\nu}$ 
values ($\gtrsim 10^{2.7}$). Over 81\% of sources with $\chi^{2}_{\nu} \gtrsim 10^{2.7}$ are 
brighter than $i =$ 17.
 
Based on the $\Sigma$ and $\chi^{2}_{\nu}$ cuts, about 73.7\% of the XRaySrc objects having 
photometric LCs are expected to be intrinsically variable. However, even if we select variable candidates 
from the XRaySrc catalog by utilizing both $\Sigma$ and $\chi^{2}_{\nu}$ together, we are unable to tell 
whether these sources are AGNs or other X-ray luminous variables solely based on these cuts. Therefore, 
a SF analysis of their optical LCs is required to identify AGNs from other types of (non-)variables. 
Furthermore, the positive correlation between these two values, seen in the SpecQSO sample at 
$\chi^{2}_{\nu} \ge$ 3, is not found in the XRaySrc sample. This is, again, likely due to both faint objects 
with large uncertainties and contamination by other sources (e.g., late-type flaring stars) other than 
AGNs/QSOs in the catalog. 

\begin{figure} [tbp]
 \begin{center}
      \includegraphics[trim=-10mm 0mm 0mm 10mm, clip, height=10cm]{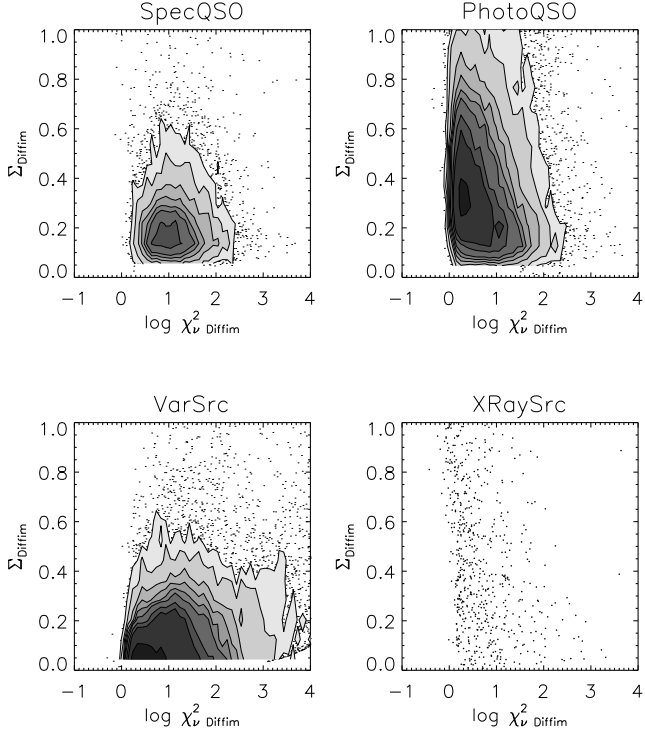}
      \caption{\label{fig8} The same as Figure~\ref{fig6} but values obtained from difference image 
      LCs.}
   \end{center}
\end{figure}

\begin{figure} [tbp]
 \begin{center}
      \includegraphics[trim=0mm 0mm 0mm 10mm, clip, height=10cm]{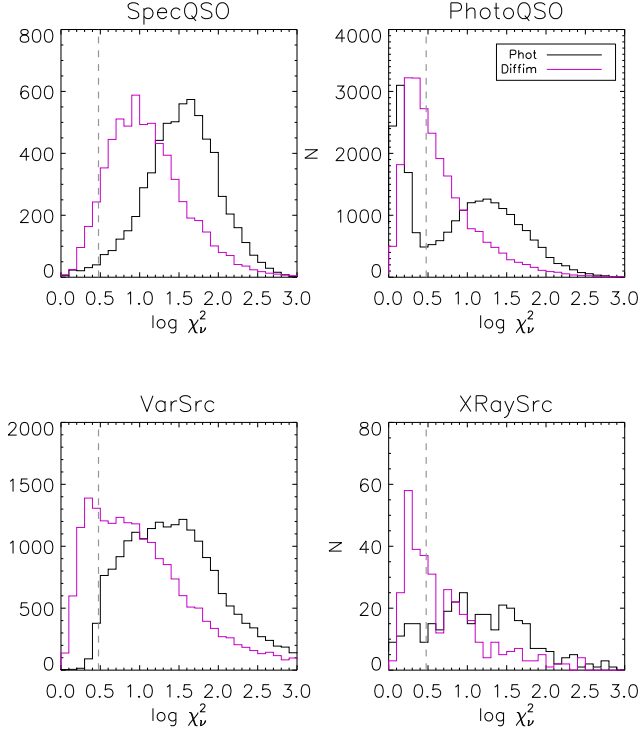}
      \caption{\label{fig9} Distributions of $\chi^{2}_{\nu}$ for sources having both photometric and
      difference image LCs. Black line is for photometric and magenta line is for difference image LCs. 
      The dashed line shows $\chi^{2}_{\nu}$ = 3. Differences between distributions of $\chi^{2}_{\nu}$ 
      mostly result from low S/N in difference image LCs.} 
   \end{center}
\end{figure}

Figure~\ref{fig8} shows the distributions of the $\Sigma$ and $\chi^{2}_{\nu}$ values for the same types 
of objects as in Figure~\ref{fig7}, but derived from difference image LCs. The lack of clear bifurcation in the 
$\chi^{2}_{\nu}$ distribution for the PhotoQSO sample, which is seen in Figure~\ref{fig7}, is due to the 
relatively large uncertainties on the difference image aperture photometry. 
We compare the distributions of $\chi^{2}_{\nu}$ measured from photometric and difference
image LCs in Figure~\ref{fig9}. In general, the distribution of $\chi^{2}_{\nu}$ from difference image 
LCs is shifted to smaller $\chi^{2}_{\nu}$ values compared to the photometric cases. The main 
reason for this is also the lower S/N in flux measurements from difference image aperture 
photometry. We discuss this further in Section~\ref{sec:diffimAdvLimSec}.  

%-------------- Fig 8 -----------------

%-------------- Fig 9 -----------------

\subsection{Structure Function For the Damped Random Walk Model}
\label{sec:DRWfittingSec}
Stochastic optical variability is believed to be intrinsic to quasars. The amplitude of 
variability is greater on longer timescales, from months to years. Although several physical
mechanisms, including accretion disk instabilities, starbursts, and microlensing, have been 
suggested, a physical origin for the quasar variability still remains an open question. 
\citet{kelly09} adopted a first order continuous autoregressive process, the damped random walk 
(DRW), to statistically model the observed light curves of quasars, and it has been shown that the DRW
model can successfully describe the observed light curves of individual quasars using large samples 
\citep{kozlowski10, macleod10}. The DRW model is a stochastic process with an exponential covariance 
\begin{eqnarray}
S_{ij}\,(\Delta\,t_{ij}) \equiv \frac{1}{2}\,\hat{\sigma}^{2}\tau\,\textrm{exp}({-\,|\Delta\,t_{ij}|\,/\,\tau}\,),
\end{eqnarray}
where $\hat{\sigma}$ controls the standard deviation of short-term variation, and $\tau$ is a characteristic
damping timescale in days. To express the long-term variability of quasars characterized by the DRW 
as a function of the time lag, \citet{macleod10} used the first-order SF,  defined as
\begin{eqnarray}  
SF(\Delta\,t) \equiv SF_{\infty}\,(1 - e^{-\,|\Delta\,t|\,/\,\tau})^{1/2}\label{drwSF}.
\end{eqnarray}
The SF reaches an asymptotic value SF$_{\infty}$\,=\,$\hat{\sigma}\sqrt{\tau}$ at large time lags, 
$|\Delta\,t| \gg \tau$. At small $|\Delta\,t|$, SF($\Delta\,t \ll \tau$)$\,=\,\hat{\sigma}\sqrt{|\Delta\,t|}$. 
These SF parameters have been found to be closely correlated with physical parameters of quasars 
such as black hole mass and luminosity \citep[e.g.,][]{macleod10, macleod12}. 

\begin{figure} [tbp]
 \begin{center}
      \includegraphics[width=8cm]{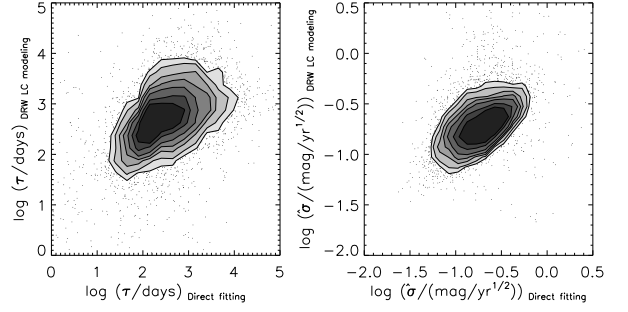}
      \vspace{-5cm}
      \caption{\label{fig10} Comparison of the $\tau$ (left panel) and $\hat{\sigma}$ (right panel) derived 
      from the two methods (direct fitting vs. DRW LC modeling) for the SpecQSO photometric LCs.}
   \end{center}
\end{figure}

\begin{figure} [tbp]
 \begin{center}
      \includegraphics[trim=10mm 0mm 0mm 10mm, clip, height=10.5cm]{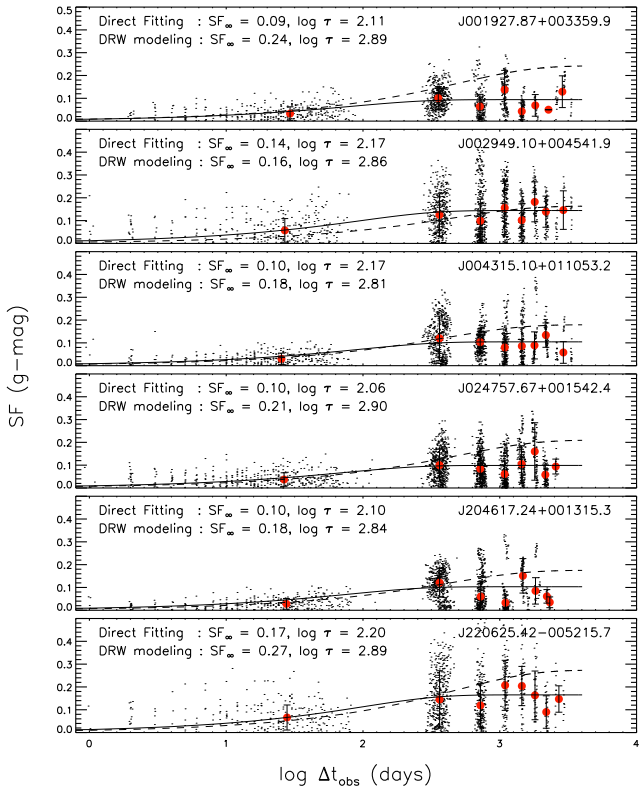}
      \caption{\label{fig11} Examples of empirical SFs for some SpecQSO objects. Black dots 
      represent the observed SF and red dots are median SF values in each time lag bin. 
      The solid (dashed) line shows the best-fit obtained by the direct fitting method (DRW LC modeling).
      Note that we are simultaneously fitting the entire time series instead of the binned values.}
   \end{center}
\end{figure}

In this work, we use two different methods to obtain the best-fit SF parameters for individual objects:(i) 
estimating $\tau$ and SF$_{\infty}$ by directly fitting the exponential form of SF (Equation~\ref{drwSF}) 
to observed SFs (Equation~\ref{obsSF}) using a least square fit, and (ii) estimating the $\tau$ and 
$\hat{\sigma}$ by modeling individual light curves using the method of \citet{kozlowski10}. 
Their method to estimate these parameters is based on the approach of \citet{press92} \citep[see the 
Appendix of][]{kozlowski10}. Note that these two methods constrain the 
amplitude of variability on differing timescales: the former method is more sensitive to long term
parameter (SF$_{\infty}$), while the latter method constrains the short term amplitude ($\hat{\sigma}$) 
better. This, and the intrinsic differences in the fitting procedure between these two algorithms result in a 
systematic difference between their best-fit SF parameters (see Figure~\ref{fig10}), although the results are 
in agreement in general. 
Since the former gives more weight to the larger time lag regime where most data points reside 
in observed SFs, it naturally returns smaller SF$_{\infty}$ and shorter $\tau$ values compared 
to those derived from the latter. One can easily see this effect in examples of observed SFs for 
some SpecQSO objects in Figure~\ref{fig11}.  

%-------------- Fig 10 -----------------

To be able to compare difference image results with the previous analyses, which were conducted in magnitude, 
we test if the same characteristic timescales are obtained for a given LC analyzed both in magnitudes and in 
fluxes by a given LC fitting methodology. The test results show that the both methods are barely sensitive to 
the linear versus log scales. Further details on the test are described in the Appendix.

%-------------- Fig 11 -----------------

\subsubsection{Outliers and Uncertainty Estimation}
\label{sec:outlierSec}
Dealing with LC outliers in variability metrics always has been problematic because they are not trivial 
to identify. One simple way to omit outliers from a light curve is to use deviation from the mean 
brightness. For example, one can define the most outlying point as an outlier or exclude data points which
are more than 3$\sigma$ from the LC mean. However, it is often difficult to tell whether they are genuine 
outliers, especially for variable sources. Visual inspection can help in removing outliers from an individual 
LC, but this will be increasingly difficult for very large sample sizes. It also can cause human bias. 

To check for the outlier sensitivity (i.e., LC robustness) and estimate the standard errors of the best fit 
parameters, we perform jackknife resampling for both methods rather than rejecting outliers based on arbitrary 
criteria. For each source, we repeat the fitting process every time after each data point in the LC is 
removed. When there are N data points in a LC, we can have N best-fits for each measured parameter, 
and then compute the rms and the median of the SF parameters from N best-fits. In some cases, we note 
that both methods are sensitive to adding or omitting even just one data point in a LC regardless of the total 
number of measurements. 
In addition to cases where LCs have actual outliers, we also find some cases where the SF parameter 
estimation largely depends on just one data point due to insufficient light curve lengths.  
Therefore, to only use well constrained (i.e., robust) LCs, we exclude sources (32\% of the SpecQSO) 
if they have $\tau$ of about zero (the ``noise'' limit) or infinity (the ``run-away'' limit) at least 
once among N best-fits of $\tau$, or have a rms in log $\tau \ge$ 0.15. Corresponding to these 
conservative robustness cuts, the fractions of robust LCs in the PhotoQSO, VarSrc, and XRaySrc catalogs 
are 46\%, 35\%, and 34\%, respectively. The fraction seems to decrease as either the contamination by 
other sources increases or the number of faint sources increases in the catalog. 
  
\begin{figure*} [tbp]
 \begin{center}
      \includegraphics[width=15cm]{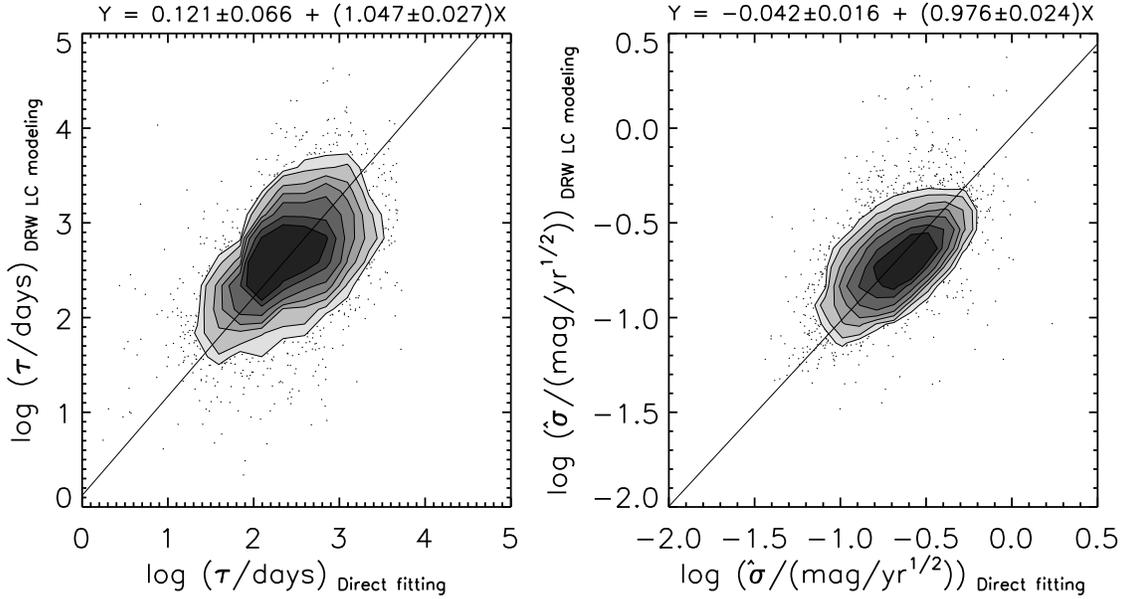}
      \vspace{-9cm}
      \caption{\label{fig12} Left panel: Comparison of the $\tau$ for robust photometric LCs (log $\tau <$ 5 
      and rms of log $\tau \le$ 0.15) of the SpecQSO sample measured by the two different methods. 
      The best-fit linear regression lines is overplotted. 
      Right panel: Same as in the left panel, but comparison of the $\hat{\sigma}$.}
   \end{center}
\end{figure*}

In Figure~\ref{fig12}, we compare log $\tau$ values for well-constrained photometric LCs of the SpecQSO 
sample, estimated from the two fitting methods. They are well related to each other, with a slope of about unity. 
We also find that most excluded sources show a broader correlation and tend to have either long characteristic 
damping timescales, log $\tau >$ 3, which is close to the maximum time lag of Stripe 82 data, or very short
damping timescales, log $\tau <$ 0. 
The non-zero intercept likely results from fundamental differences in the fitting procedure 
between the two methods mentioned above, which can lead to systematic differences in estimated $\tau$. 
Figure~\ref{fig12} also shows the comparison of best-fit $\hat{\sigma}$ for well-constrained LCs. 
There is good agreement between results from the two methods. 

In this section, we have shown that the simple direct fitting method is sufficient enough to estimate SF 
parameters. Systematic differences from the DRW LC modeling can be taken into account when 
identifying AGNs in the SF parameter space (Section~\ref{sec:identifyAGNByVarSec}). 
We have also shown that the direct fitting method is insensitive to the scales of a LC (flux vs. magnitude). 
Therefore, we will focus on the best-fit SF parameters estimated by the simple direct fitting method, which is 
computationally less intensive, in our further analysis.

%-------------- Fig 12 -----------------

\subsection{Comparison of Photometric and Difference Image Light Curves}
\label{sec:CompOfPhotoDiffimLCsSec}
Insensitivity of the direct fitting method to the LC scales allows SF analysis on difference image LCs, which are 
solely in flux, and a direct comparison of photometric LCs to {\it common-epoch} difference image LCs in 
terms of the best-fit $\tau$ values to see if there are any differences. We then compare photometric LCs 
to {\it all-epoch} difference image LCs to explore the effect of number of photometric measurements on 
the SF parameter estimation. Again, all LCs used in this section are from sources in the SpecQSO catalog. 

First, we construct a set of difference image LCs having exactly the same-epoch data points as
their counterpart photometric LCs. The common epoch requirement reduces the sample size. 
However, this strict condition ensures that we can detect any possible differences between the results of the 
SF analysis from photometric and difference image LCs, mainly caused by their different photometry 
(PSF vs. aperture), and thus, different S/N. Using the direct fitting method, we estimate the SF parameters 
for the set of common-epoch difference image LCs. 
For the comparison of log $\tau$ values between photometric and difference image LCs, we have 
the slope of 1.006$\pm$0.167 and the y-intercept of 0.271$\pm$0.339 (larger log $\tau$ for 
photometric LCs). Here, the large uncertainties are likely due to the small sample size. A non-zero intercept 
indicates that the lower S/N of difference image LCs, which merely resulted from aperture 
photometry, can affect the estimation of $\tau$ in a way that systematically decreases it. This implies 
that our current difference image LCs might be less sensitive to low levels of variability, which is usually 
linked to short damping timescales, than photometric LCs. We repeat the same test with the DRW LC
modeling and find the same trend. We expect that this will be improved by doing PSF photometry on the 
difference images generated from more precise PSF-matched template images in future works 
(e.g., https://dev.lsstcorp.org/trac/wiki/DC/Winter2013).

However, ideally we should be able to constrain variability of AGNs better with a difference image LC since 
it has, on average, $\sim$20 more data points than a photometric LC. To confirm 
this, we use the all-epoch difference image LCs of the same objects in the common-epoch LC set. Firstly, 
with all-epoch difference image LCs, we can estimate the SF parameters for about 14\% more objects that 
were rejected in the common-epoch LC set due to lack of number of measurements (N $<$ 4). Secondly, 
more objects (about 6\% of the sample) pass the LC robustness cuts while they were described as 
unconstrained with the common-epoch LCs (i.e., the rms of log $\tau \ge$ 0.15, the run-away limit, or 
the noise limit). Thus, we conclude that more data points in a LC can definitely help in constraining AGN 
variability by increasing number of objects passing the initial N$_{obs}$ cut and improving the LC robustness. 
We expect that this somewhat compensates for the disadvantage of the lower S/N in difference image LCs 
for now, and eventually leads to significant improvements in the context of AGN identification in difference 
images with PSF photometry in the future. We will discuss advantages and limitations of difference image LCs 
in more detail in the following section.

\subsection{Advantages and Limitations of Difference Image Light Curves}
\label{sec:diffimAdvLimSec}
Variability-based AGN selection methods have been limited to mostly point-like sources in the optical. 
This is because light contamination by extended host galaxies may dilute the change in brightness arising 
from relatively weak accretion activity, and the lower variability amplitudes result in difficulties in identifying 
them as AGNs. Thus image differencing works best in selecting AGNs embedded in extended host galaxies 
or lower-luminosity AGNs since it does not need to assume or fit a source model, and only the variable nuclei  
will stand out in difference images. Difference image LCs that are free of host galaxy contamination enable 
us to apply the variability-based selection to host-dominant AGNs as well. We will explore the variability 
properties of X-ray detected optically-extended sources, which are candidate faint AGNs, in Section~\ref{sec:resolvedSrcSec}.

\begin{figure} [tbp]
 \begin{center}
      \includegraphics[trim=0mm 0mm -2mm 10mm, clip, width=8.5cm]{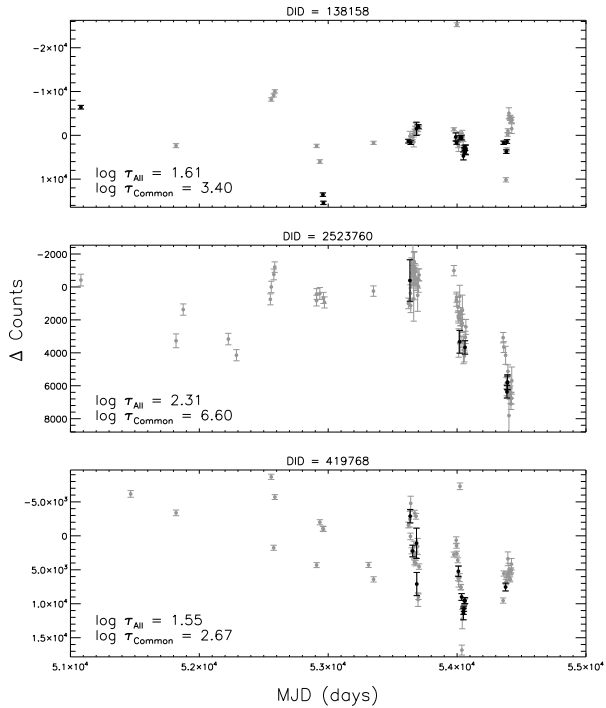}
      \caption{\label{fig13} Examples of LCs to show the effect of more data points in a LC on 
      estimating the SF parameters, especially $\tau$. Gray dots are for difference image LC data points and 
      black dots are for the common-epoch data points. Top: SDSS J004032.10-001350.8 
      Middle: SDSS J005149.24-000133.5 Bottom: SDSS J230946.14+000048.8}
   \end{center}
\end{figure}

As described in Section~\ref{sec:sDSSDiffImSec}, a difference image LC contains, on average, 
about 20 more detections compared to its counterpart photometric LC. Figure~\ref{fig13} shows 
some examples of difference image LCs that illustrate the effect of more detections on the estimation 
of log $\tau$. Even if the maximum time lag of observation is similar, missing data points between 
measurements makes a significant difference in the LC cadence (top panel of Figure~\ref{fig13}), 
leading to the different SF parameters. The length of LCs is another important factor in determining 
variability properties. \citet{macleod11} tested the impact of LC length on the best-fit distribution of 
the SF parameters and found a significant bias in log $\tau$ and SF$_{\infty}$ when estimated from 
LCs with a length significantly shorter than 10 years. They showed $\tau$ can be easily overestimated 
from short LCs. Indeed, we see the impact of the LC length on $\tau$ in the middle and bottom panels of 
Figure~\ref{fig13}. The common-epoch data cover only a small fraction of the entire temporal baseline, 
and returns very different $\tau$ values (larger than those measured from all-epoch data, even returning 
a run-away timescale for some cases). These results suggest that using LCs with more measurements 
(i.e., difference image LCs in this study) to characterize AGN variability may improve the variability-based 
AGN selection. 

%-------------- Fig 13 -----------------

\begin{figure} [tbp]
 \begin{center}
      \includegraphics[trim=10mm 5mm 0mm 20mm, clip, height=10cm]{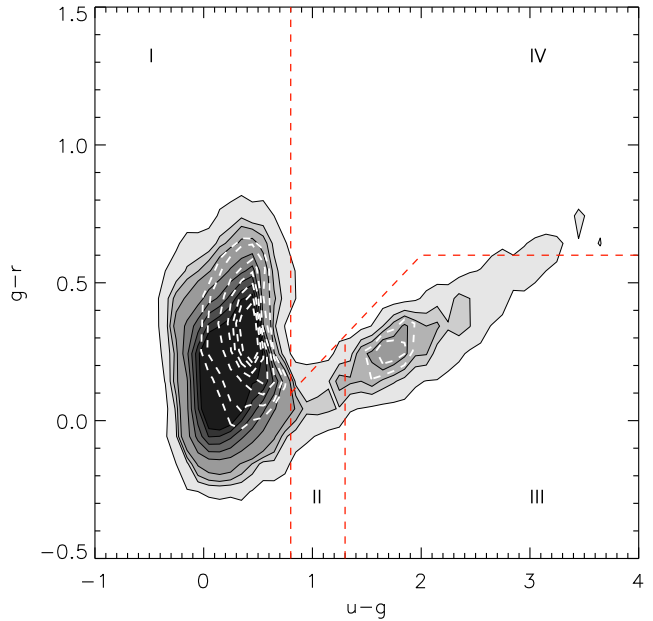}
      \vspace{-1.5cm}
      \caption{\label{fig14} $u$-$g$ vs. $g$-$r$ CCD for the PhotoQSO sample. 
      Black contours show the entire sample while white dashed contours represent sources with 
      $\chi^2_{\nu} < $3.}
   \end{center}
\end{figure}

A weakness of our difference image LCs is relatively large uncertainties on flux measurements 
that resulted from aperture photometry, while the recalibrated photometric data set \citep{ivezic07}
uses PSF magnitudes providing more precise photometric LCs at faint magnitudes. 
Several features, probably caused by lower S/N of difference image LCs, are detected. 
For example, there is a shift of the $\chi^{2}_{\nu}$ distribution to smaller values for difference image LCs, 
disappearance of the bifurcation in the $\chi^{2}_{\nu}$ distribution for the PhotoQSO sample that is seen in 
the photometric LC analysis, and systematically smaller best fit log $\tau$ values for the common-epoch 
difference image LCs. Since there is a significant change in the shape of $\chi^{2}_{\nu}$ distribution (rather 
than just a shift of the entire distribution), loosening the $\chi^{2}_{\nu}$ criterion for difference image LCs is 
not helpful to construct a clear sample of intrinsically variable sources. Instead, this will cause high contamination 
in the sample by non-variable/faint sources with variability dominated by larger photometric noise.
Figure~\ref{fig14} shows a $u-g$ vs. $g-r$ CCD for the PhotoQSO sample. Sources 
with $\chi^{2}_{\nu} < $3 occupy the elongated part in Region I on the CCD. That no such feature is seen in 
the SpecQSO or the VarSrc samples implies that these sources are mostly non-variable. Visual inspection of 
their spectra in the SDSS DR9 \citep{ahn12} indicates many of them are actual contaminants such as stars 
and galaxies or else are highly absorbed/reddened quasars. 

However, we are able to minimize the effect of large uncertainties in difference image LCs on AGN selection 
by imposing appropriate selection criteria, which take systematic differences in the SF parameters into account 
(see Section~\ref{sec:identifyAGNByVarSec}). Furthermore, we stress that the SDSS is not optimized for 
image differencing and the primary goal of this study is to lay some groundwork for new time domain surveys 
that will really focus on obtaining higher-quality difference image data. 

%-------------- Fig 14 -----------------

\subsection{Identifying QSOs by Variability Seen in Difference Images}
\label{sec:identifyAGNByVarSec}
\subsubsection{Completeness and Efficiency}
To evaluate our AGN selection based on optical variability seen in difference images and compare to 
\citet{macleod11}, we compute the completeness (the fraction of the recovered SpecQSO) for given 
selection cuts using well-constrained LCs in the SpecQSO catalog. \citet{macleod11} limit their 
sample to bright sources (i $<$ 19) when estimating the completeness and the efficiency (i.e., purity) for 
their selection cuts because the SpecQSO catalog is complete for i $<$ 19 \citep[Table 2 in][]{macleod11}. 
The completeness is defined as percentage of recovered quasars out of the entire confirmed quasar and the 
efficiency is defined as percentage of confirmed quasars out of selected objects (i.e., purity).

For bright sources, we achieve a completeness of 82.2\%, 96.4\%, and 98.6\% for log $\tau \ge$ 2, 1.5, 
and 1, respectively. The completeness for log $\tau \ge$ 2 is lower compared to \citet{macleod11}, since 
we have systematically smaller log $\tau$ values due to the differences in the methodologies of estimating 
the SF parameters (the direct fitting vs. the DRW model; the mean offset of 0.121$\pm$0.066 dex) and 
the difference in flux measurements (PSF vs. aperture photometry; the mean offset in log $\tau$ of 
0.271$\pm$0.339 dex). When considering these combined offsets, lowering the selection cut on log 
$\tau$ to $\sim$1.5 is reasonable and it allows a comparable completeness to that for log $\tau \ge$ 2 
in \citet{macleod11}, which was 94\% (93\% after omitting the most outlying data point in each LC). 
In fact, our completeness (96.4\%) for log $\tau \ge$ 1.5 is more similar to that (96\%) of 
\citet{macleod11} when they apply both log $\tau \ge$ 1.5 and $\Delta\,L_{noise} >$ 10 
(where $\Delta\,L_{noise}$ is the log likelihood of a DRW solution minus the log likelihood of a white 
noise solution). This indicates that our LC robustness cuts, especially the rms of log $\tau$ cut, functions as 
the constraint $\Delta\,L_{noise} >$ 10 in that both of them gauge the suitability of the DRW model for a 
given LC and boost the completeness and efficiency of the AGN selection.  

\begin{figure} [tbp]
 \begin{center}
      \includegraphics[trim=10mm 30mm 0mm 20mm, clip, height=9cm]{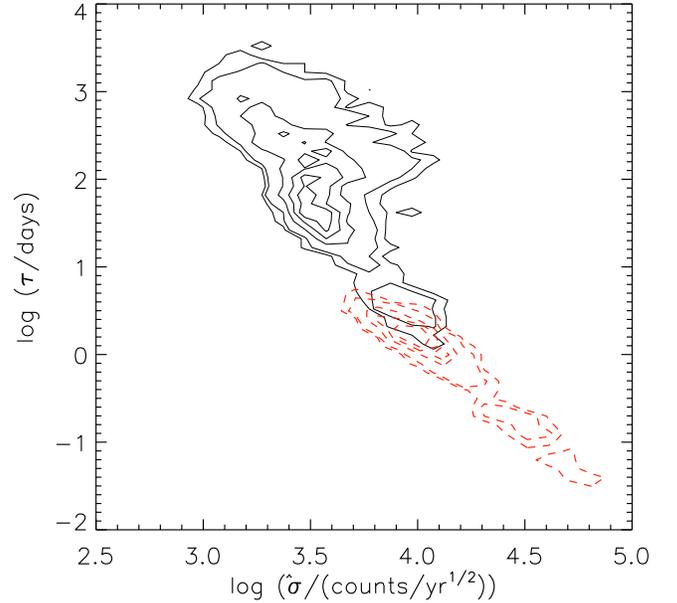}
      \caption{\label{fig15} Characteristic damping timescale $\tau$ as a function of the driving amplitude of
      short-term variation $\hat{\sigma}$ for the SpecQSO sample (black solid) and for the control sample
      (red dashed). We draw contours at 5, 10, 30, 50, and 70 data points.}
   \end{center}
\end{figure}

Figure~\ref{fig15} shows the distribution of the SF parameters for the entire SpecQSO (black solid) 
and for the control sample (red dashed), which represents either non-variable sources or empty spots 
on the sky (i.e., white noise). Including faint quasars ($i \ge$ 19) drops the completeness to 
56.7\%, 81.2\%, and 91.4\% for log $\tau \ge$ 2, 1.5, and 1, respectively. Unlike bright sources, faint 
sources in the SpecQSO can have log $\tau <$ 1.5. This is mainly because relatively large uncertainties 
due to aperture photometry make difference image LCs of fainter sources indistinguishable from those of 
non-variable sources. It leads to a shift of the distribution of fainter sources towards the non-variable region 
in the SF parameter space. To maintain a reasonably high completeness of 91.4\%, one can adopt log 
$\tau \ge$ 1 as an AGN selection criterion. This cut statistically separates quasars well from the control 
sample; over 99\% of the control sample shows log $\tau <$ 1. The completeness for various $\tau$ 
criteria are listed in Table~\ref{specQSOcomp}. 

%-------------- Fig 15 -----------------

\begin{figure} [tbp]
 \begin{center}
      \includegraphics[height=9cm]{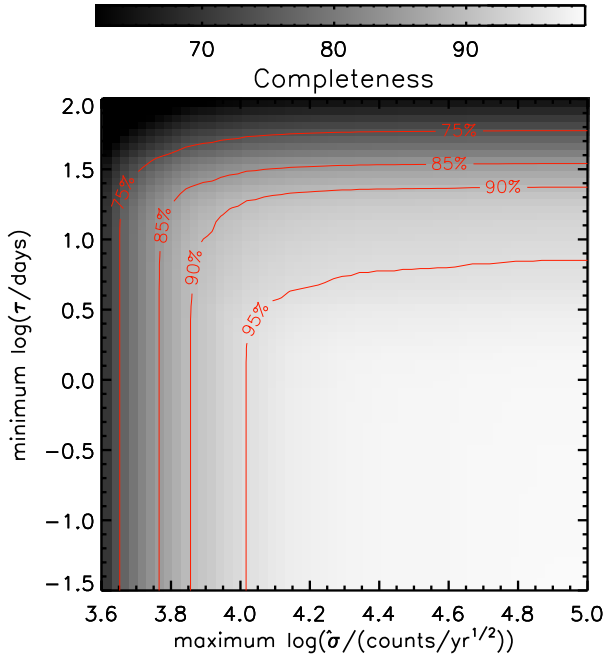}
      \caption{\label{fig16} The completeness for AGN selection criteria for the VarSrc sample.}
   \end{center}
\end{figure}

\begin{figure} [tbp]
 \begin{center}
      \includegraphics[height=9cm]{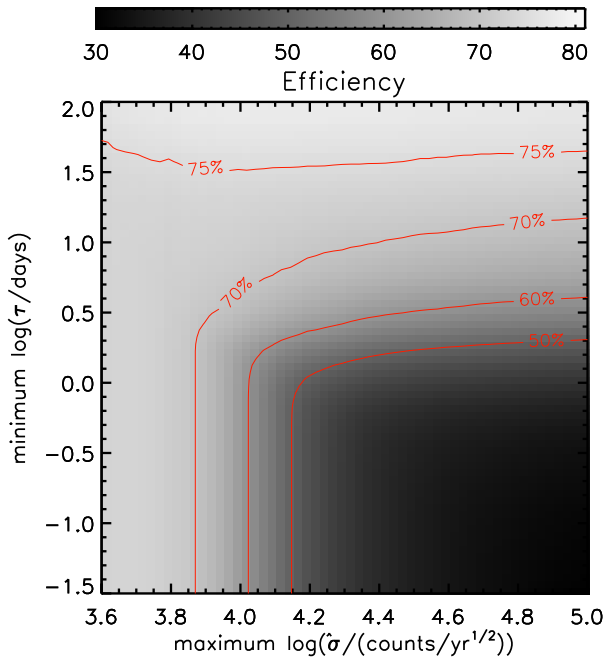}
      \caption{\label{fig17} The efficiency for AGN selection criteria for the VarSrc sample.}
   \end{center}
\end{figure}

The VarSrc catalog contains 6,573 spectroscopically confirmed quasars ($\sim$70\% of the SpecQSO 
sample). The large number of non-quasar contaminants in the VarSrc catalog allows us to assess the 
efficiency of the AGN selection based on their variability. The estimated selection efficiency here, however, 
is the lower limit due to incompleteness of the subsample of confirmed quasars in the VarSrc catalog even at 
the bright end. In terms of brightness, this subsample is between the bright and the entire SpecQSO samples 
given the cut at $g <$ 20.5. Thus, the completeness of this subsample for the same selection cut is also 
expected to be between those for the bright and the entire SpecQSO samples. The completeness and 
efficiency as a function of the minimum log $\tau$ and maximum log $\hat{\sigma}$ are shown in 
Figure~\ref{fig16} and Figure~\ref{fig17}, respectively. Completeness of 65\% and efficiency of 78\% 
can be achieved for log $\tau \ge$ 2 and log $\hat{\sigma} \le$ 4.15. If we decrease the minimum log 
$\tau$ cut to 1.5 while maintaining the cut on log $\hat{\sigma}$, the completeness significantly rises to 
86\% while the efficiency (75\%) decreases only by a few percent. By lowering the minimum $\tau$ 
criterion, one can select many more quasars without adding many contaminants. When reducing the 
minimum log $\tau$ cut further down to 1, the completeness can increase to 93\% with the efficiency of 
71\% and this selection criteria gives the maximum $\sqrt{C^2+E^2}$. The results are summarized in 
Table~\ref{varcomp}. 

We explore contaminants in detail to see if there are possible quasars that were not 
spectroscopically confirmed yet in the SDSS DR7 data set. We first select sources with log $\tau \ge$ 1.5, 
which is pretty conservative cut, that are not found in the SpecQSO catalog. Among 940 selected sources, 
733 have colors consistent with low-redshift quasars (Region I on the $u$-$g$ vs. $g$-$r$ CCD ). 
We could retrieve 208 spectra from the DR9 data set and 4 spectra from spectroscopic follow-up with 
Apache Point Observatory DIS \citep{chelseaPhD}, and find that 56\% of them are confirmed as quasars.
If we assume that the fraction of quasars (56\%) remains the same, then the efficiency for the log 
$\tau \ge$ 1.5 cut is expected to increase by at least 10\% for a spectroscopically complete sample. 

When adopting log $\tau \ge$ 1 (1.5, 2) as an AGN selection criterion, 66\% (52\%, 30\%) of the 
PhotoQSO sample are classified as AGNs. A large fraction of sources ($\sim$40\%) with log $\tau <$ 1 
show photometric $\chi^{2}_{\nu} <$ 3, indicative of observed variability dominated by photometric noise. 
These sources account for about 13\% of the entire PhotoQSO sample. Thus, we can conclude that 
there are at least $\sim$13\% of possible contaminants in the PhotoQSO based on our variability analysis. 
There are a few sources (21 objects) whose characteristic damping timescales are in the range of 1 to 10 
days and that are bright in the X-ray. They actually show significant evidence for intrinsic optical variability 
(i.e., $\chi^{2}_{\nu} \ge$ 3 and $\Sigma \ge$ 0.05),  but are not found in the SpecQSO catalog. Some of 
them could be faint typical AGNs with relatively short damping timescale. Another possible explanation 
for these objects is that they might be blazars.

%-------------- Fig 16 -----------------

%-------------- Fig 17 -----------------

\subsubsection{Blazar Candidates}
Our discussion of AGN variability thus far has focused primarily on the optical variability characteristics of 
the thermal disk continuum emission from typical Type I quasars. However, some rare AGNs (blazars) are 
instead dominated in their optical emission by a relativistic jet aligned with the line of sight \citep{blandford78}; 
these jet-dominated objects are known to be even more variable than typical quasars due to the effects of 
relativistic beaming. \citet{ruan12} have shown that blazar optical light curves are also well-described as a 
DRW process, but with smaller $\tau$ and larger $\hat{\sigma}$ than typical quasars. These blazars should 
thus stand out as extreme variability outliers in our difference image LCs, allowing for variability-based 
identification. As a test of our ability to detect this non-thermal jet emission through optical variability in our 
difference image LCs, we select extreme variability objects with log $\hat{\sigma} >$ 4.2 and 
1 $<$ log $\tau <$ 3 in the VarSrc sample. These cuts will select objects with smaller $\tau$ and larger 
$\hat{\sigma}$ than the vast majority of confirmed quasars (e.g., in the SpecQSO sample in 
Figure~\ref{fig15}), and are consistent with emission affected by relativistic beaming in an aligned jet 
\citep{ruan12}.

\begin{figure*} [tbp]
 \begin{center}
      \includegraphics[trim=0mm 70mm 0mm 75mm, clip, width=18cm]{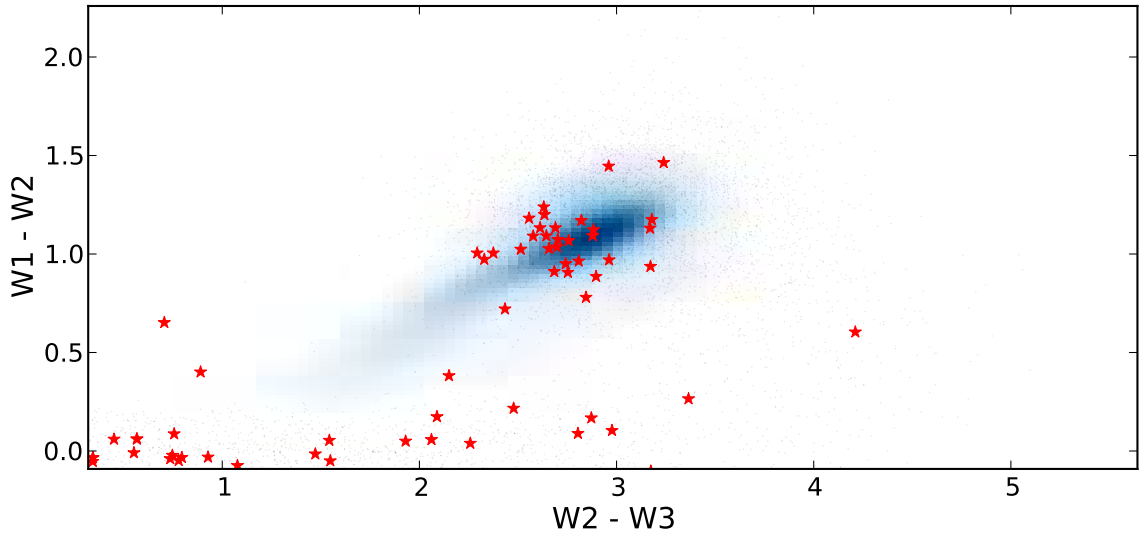}
      \caption{\label{fig18} WISE W1-W2 vs. W2-W3 CCD of all VarSrc sources passing the robustness cuts, 
      matched to WISE sources with S/N $>$ 3 in the WISE W1 [3.4 $\mu$m], W2 [4.6 $\mu$m], 
      and W3 [12 $\mu$m] bands (black points). Highly variable sources with log $\hat{\sigma} >$ 4.2 and 
      1 $<$ log $\tau <$ 3 are highlighted as red star symbols. A Gaussian Kernel Density Estimate of the 
      location of the WISE blazar strip is shown in blue. Sources with W1 - W2 $<$ 0.5 tend to be stars and 
      galaxies.}
   \end{center}
\end{figure*}

\citet{massaro11} showed that blazars detected in Wide-field Infrared Survey Explorer \citep[WISE;][]{wright10} all-sky survey tend to lie along a narrow `WISE blazar strip' on WISE CCD. 
To gauge whether our highly-variable objects are of blazar origin, we positionally match the VarSrc sources to 
the WISE all-sky catalog, using a 3$\arcsec$ matching radius. Figure~\ref{fig18} shows the WISE W1-W2 
vs. W2-W3 filter CCD of these matches that have S/N $>$ 3 in the WISE W1 [3.4 $\mu$m], 
W2 [4.6 $\mu$m], and W3 [12 $\mu$m] bands. Highly variable sources in this sample 
(log $\hat{\sigma} >$ 4.2 and 1 $<$ log $\tau$ $<$ 3) are highlighted as red star symbols. 
We also show the WISE blazar strip by overplotting the results of a Gaussian Kernel Density Estimate using 
the 3,021 known blazars in the Roma-BZCAT blazar catalog (Massaro et al. 2009) which matched to WISE 
sources with S/N $>$ 3 in W1, W2, and W3. In Figure~\ref{fig18}, the cloud of variable sources (black points 
approximately centered at (W1-W2, W2-W3) $=$ (1.1,  3.1) are mostly AGN, while sources with W1-W2 
$<$ 0.5 tend to be stars and galaxies. Many of the highly variable sources highlighted appear to be AGN, as 
well as many stars. Although the WISE blazar strip is relatively narrow, much of it lies directly on top of the 
cloud of typical (non-jet dominated) AGNs, making it difficult to determine whether a source near the WISE 
blazar strip is dominated by non-thermal jet emission. The relatively low efficiency of blazar selection using the 
WISE blazar strip (especially for the flat-spectrum radio quasar sub-class of blazars) has been noted previously 
\citep{ruan12}. Nevertheless, it is intriguing that the positions of the highly-variable sources with W1-W2 
$\ge$ 0.5 do not appear to sample the distribution of the more numerous quasars in the background, but 
rather seem to lie preferentially closer to the WISE blazar strip. Thus, it is possible that many of these sources 
are blazars, with blazar-like WISE colors and extreme optical variability. In fact, one third of them are found in 
the Stripe 82 radio catalog by \citet{hodge11}. While it is clear that WISE colors alone do not efficiently 
separate blazars from non-jet dominated quasars due to their strong overlap in the WISE CCD, our inclusion 
of additional variability information can greatly help in narrowing down the candidates. 

%-------------- Fig 18 -----------------

\subsection{Identifying AGNs in X-Rays}
\label{sec:identifyAGNByXRaysSec}
\begin{figure} [tbp]
 \begin{center}
      \includegraphics[height=10cm]{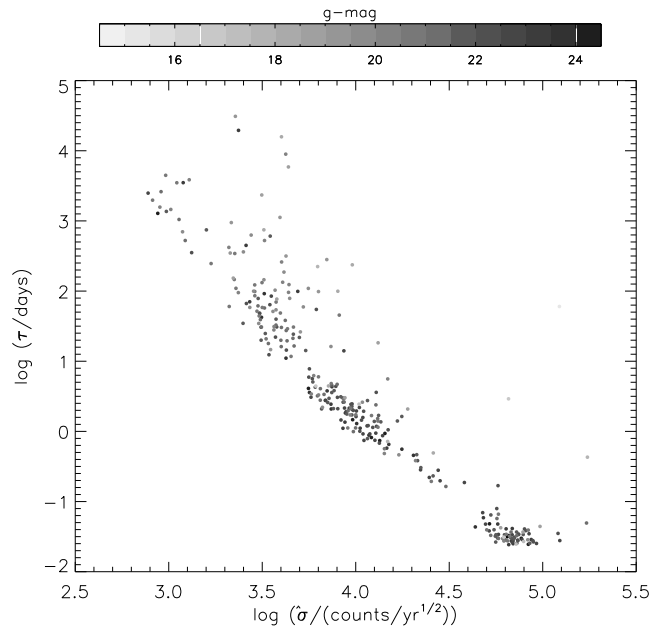}
      \vspace{-1cm}
      \caption{\label{fig19} Distribution of the SF parameters for X-ray selected point-like sources, 
      color coded by $g$ magnitude. Faint sources tend to have shorter characteristic timescales. Although large 
      and uncorrelated photometric noise may spuriously induce apparent short timescale variability in optically 
      faint objects, X-ray detection provides additional power to efficiently discriminate true faint AGNs from 
      contaminants (see discussion in Section~\ref{sec:identifyAGNByXRaysSec}).}
   \end{center}
\end{figure}

Figure~\ref{fig19} shows the distribution of the SF parameters for X-ray selected optically-unresolved 
sources before applying the cut on rms of log $\tau$. The data are color coded by their magnitude in 
$g$-band. Most sources that are fainter than $g \simeq$ 21.5 tend to have short 
damping timescales (log $\tau <$ 1). Some of them are intrinsically faint in the optical, and some of 
them are faint due to obscuration (e.g., Type II AGN). A strong magnitude dependence of the 
completeness of variability-based AGN selection is well-known \citep[e.g.,][]{kozlowski10}. Given that 
photometric noise becomes large and uncorrelated between observations for faint sources, reported small 
$\tau$ values for faint AGNs may be spuriously induced. If AGNs are not bright enough in the optical, their 
optical LCs could be neither recognized as typical AGNs nor well constrained with DRW model (i.e., spurious  
small $\tau$ with rms of log $\tau >$ 0.15). This implies that applying the same AGN selection cuts regardless 
of magnitude can result in missing many faint AGN candidates. According to visual inspection of spectra 
of objects with log $\tau \ge$ -1, 95\% are indeed classified as QSOs (only a small fraction of contamination 
by active late-type stars). A much higher contamination rate is found in sources with log 
$\tau <$ -1. Therefore, we conclude that X-ray detection provides additional power to efficiently discriminate 
true faint AGNs from the contaminants even down to log $\tau \ge$ -1, the regime where it is usually 
impossible to separate true faint AGNs from objects dominated by photometric noise solely based on the 
variability properties. Note that we do not claim that these short $\tau$ values are related to physical 
timescales of any accretion processes. Even though photometric noise for faint sources will be improved by 
deep imaging surveys and accurate PSF photometry, X-ray detection as an additional constraint would still be 
very helpful to finding faint AGNs, and thus boosting the selection completeness and efficiency. 

%-------------- Fig 19 -----------------

To probe the origin of AGN optical variability, correlation between optical/UV and X-ray variability has been 
studied in terms of both time lag and relative variability amplitude between two bands \citep[e.g.,][]{clavel92, 
peterson00, edelson00, nandra00, shemmer03, uttley03, arevalo09, kelly11, cameron12}. However, there 
has been debate over the nature of optical variability since some AGNs (e.g., NGC 5548 or NGC 4395) show 
a strong correlation, with a high cross-correlation coefficient, between the two bands on both short and 
long-term timescales whereas some show no clear correlation at all (e.g., NGC3516 or NGC 7469). 
In some other AGNs, the two bands are uncorrelated on shorter timescales while they are correlated on longer 
timescales, and vice versa (e.g., NGC 4051 or Ark 564). This puzzling variety in observed relations between 
optical/UV and X-ray variability implies that the process is too complex to be explained solely by a single 
model even in a single AGN, such as reprocessing of X-rays \citep{guilbert88}, propagating of intrinsic random 
fluctuation in the accretion rate inward through accretion flow \citep{arevalo08} and Compton up-scattering of 
optical/UV photons \citep{haardt91}. Lack of large samples that are simultaneously observed in both bands 
with high cadence over long timespans have made it hard to understand their connection. Alternatively, it may 
suggest that an AGN accretion disk fundamentally requires all combined effects of these mechanisms at the 
same time. 

\begin{figure} [tbp]
 \begin{center}
      \includegraphics[height=10cm]{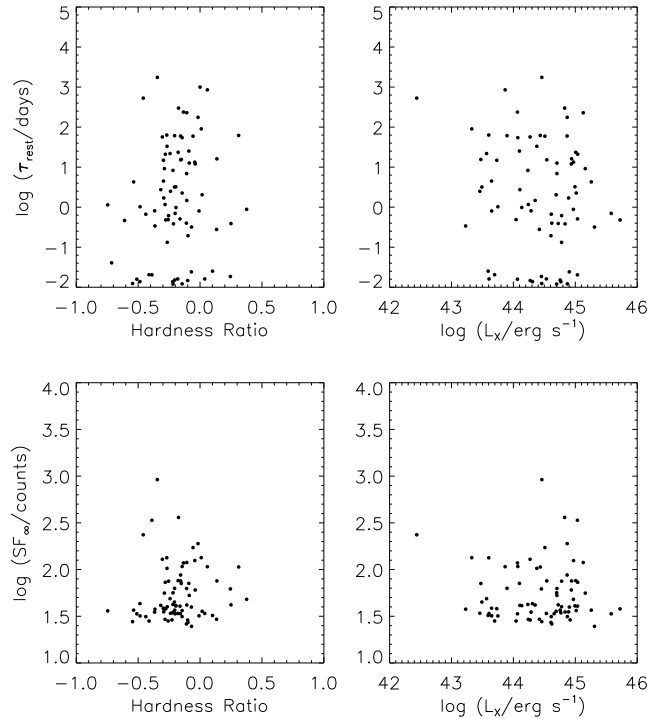}
      \vspace{0.5cm}
      \caption{\label{fig20} Comparison between the rest-frame SF parameters and X-ray properties.}
  \end{center}
\end{figure}

We search for correlations between the optical variability and X-ray properties, such as spectral shape and 
luminosity. 
The hardness ratio, (F$_{Hard}$-F$_{Soft}$)/(F$_{Soft}$+F$_{Medium}$+F$_{Hard}$), is used as 
an approximation of the X-ray spectral shape \citep{evans10}. Here, F denotes the aperture photon flux in 
each X-ray band (soft:0.5-1.2~keV, medium:1.2-2.0~keV, and hard:2.0-7.0~keV). For objects having 
redshift information in DR9, we compare the optical variability to the X-ray properties in the rest frame in 
Figure~\ref{fig20}. We find no correlations between hardness ratio and either SF$_{\infty}$ or $\tau$ 
(top panels). No noticeable correlations between the optical variability and hardness ratio 
indicate that the X-ray spectral shape might be associated only with the rapid variability in X-ray 
\citep{konig97}, which would not be well reflected in optical LCs if we consider the reprocessing mechanism. 
The correlations, if they indeed exist, also could be washed out because the optical variability properties are 
drawn from multi-epoch observations while the hardness ratio is measured from a single-epoch observation. 
\citet{grupe10} studied 92 bright soft X-ray selected AGNs and they do not detect significant level of 
X-ray spectral variability on timescales of a year or two, indicative of no correlation with long-term optical 
variability. However, \citet{nandra00} find that the X-ray spectral shape is positively correlated with the UV 
flux in NGC 7469, supporting Comptonization, although UV and X-ray fluxes are uncorrelated.  We do not see 
any statistically meaningful correlations between the optical variability and X-ray luminosity, thus no trend that 
X-ray luminous AGNs are more variable \citep{macleod10, butler11, young12}.

%-------------- Fig 20 -----------------

\section{VARIABILITY OF RESOLVED SOURCES}
\label{sec:resolvedSrcSec}
\subsection{The Optical Variability and Spectral Properties}
To describe the whole population of AGNs and its evolution over cosmic time, compiling a sample of extended
(i.e., host-dominated) AGNs, which are usually omitted from photometric or even from spectroscopic catalogs 
of QSOs/AGNs, is necessary. While current optical variability surveys focus on unresolved bright sources, 
image differencing will be the fundamental means of distinguishing variability in {\it extended} sources for 
upcoming surveys. Since image differencing only leaves signal from variable regions, we are able 
to identify AGN activity blended with extended host galaxies, where traditional photometric selection is 
challenged. Close pairs of AGN also can be identified without fiber collisions that complicate multi-object 
spectroscopic searches. 

Previous optical variability studies of extended sources have been accomplished using high angular resolution 
images to minimize the effect of the host galaxy light on the nuclear variability \citep[e.g.,][]{bershady98, 
klesman07, trevese08, boutsia09, cameron12}. However, it was impossible to perfectly remove light 
contamination by surroundings from flux measurement of the nuclear activity. Furthermore, most previous 
studies utilized imaging data monitored for less than 2 years with small coverage of the sky. 

Our final goal in this paper is to explore optical properties of X-ray detected optically--extended sources, and 
to evaluate variability selection as a tool for distinguishing AGNs from star-forming galaxies and other types of 
extended sources that can contaminate AGN catalogs. To do this, we first construct a subsample of 537 
candidates of extended AGN from the XRaySrc catalog by removing objects that are unresolved in the optical 
when the optical information is available. Since most of X-ray detected AGN candidates are faint in the optical, 
only 222 objects (41\% of the subsample) have optical counterparts. More than half of them are fainter than 
$g =$ 21.5, and their magnitude distribution is similar to that of X-ray detected unresolved sources. Based on 
their optical variability, $\sim$67\% of them can be selected as AGN candidates (log $\tau \ge$ -1; see 
Section~\ref{sec:identifyAGNByXRaysSec}). With a more conservative cut, log $\tau \ge$ 0, half of them 
can be selected as AGN candidates. None of them are suspected as lensed QSO candidates \citep{lacki09}.

%-------------- Fig 21 -----------------
\begin{figure*} [tbp]
 \begin{center}
      \includegraphics[trim=15mm 145mm 20mm 10mm, clip, width=20cm]{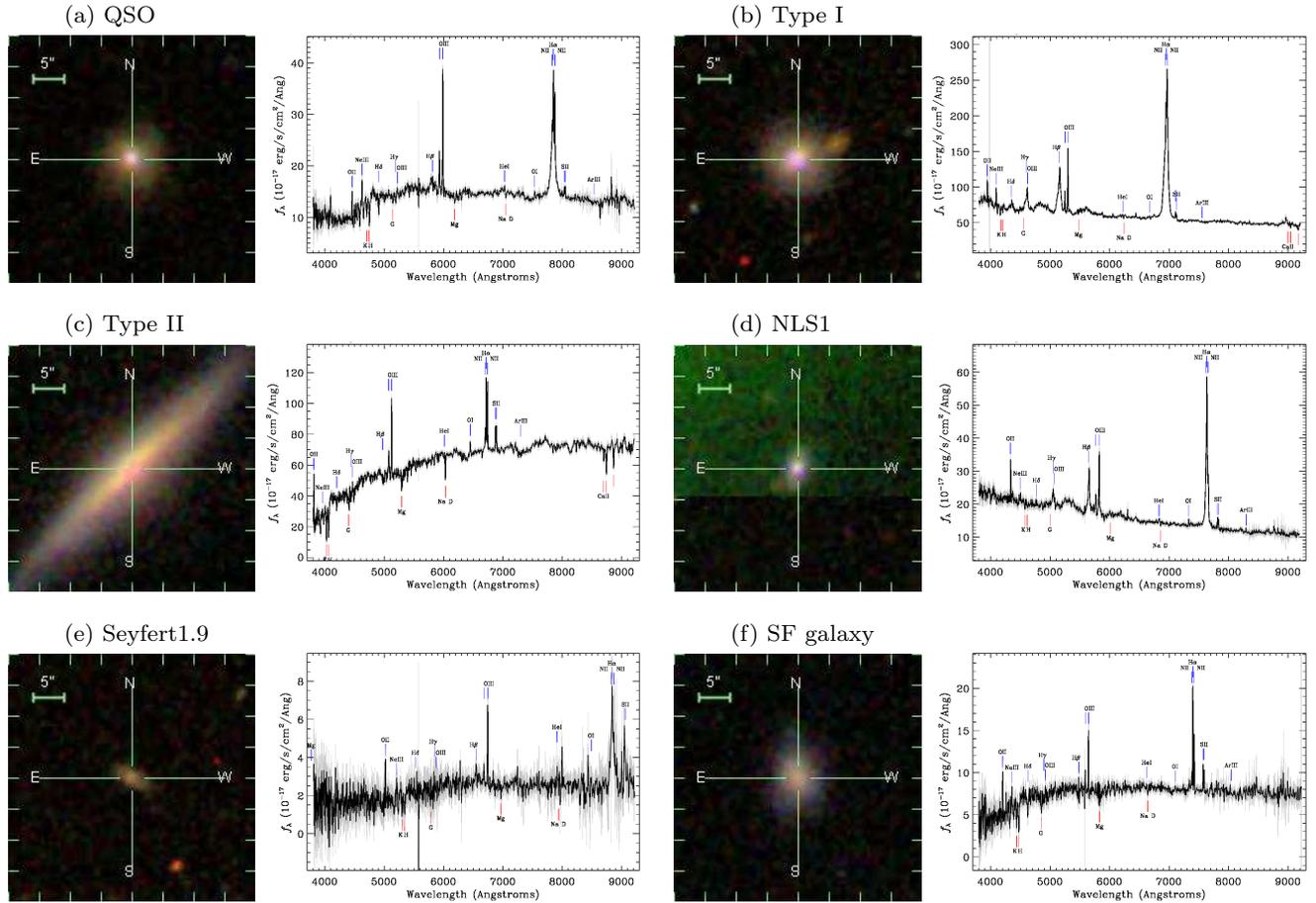}
      \caption{\label{fig21} Examples of images and spectra for some X-ray detected optically-extended 
      sources.}
  \end{center}
\end{figure*}

We investigate available optical spectra of 89 (out of 222) optically--extended AGN candidates by simple 
visual inspection. The majority of them show clear emission lines. According to their spectral properties, such as 
line ratio and width, the sample of 89 AGN candidates consists of various types of AGNs (66\%) 
as well as star-forming galaxies (11\%), other emission-line galaxies (6\%), X-ray bright optically inactive 
galaxy (XBONG) candidates (6\%), composite galaxies (4\%), normal galaxies (4\%), and unknown objects 
(3\%). Among AGNs, we further divide their spectral class in 5 sub-classes: QSO, Type I/II, narrow-line 
Seyfert1 (NLS1), and Seyfert 1.9. Classification is performed based on the traditional definitions of each 
sub-class. In Type II spectra, only the narrow lines are present. NLS1 galaxies are known to have smaller 
black hole masses that are accreting very close to the Eddington rate, thus bright in the X-ray, and exhibit lower 
amplitudes of optical variability \citep{ai13}. In NLS1, a strong FeII line is expected. NLS1 is also characterized 
by weak [OIII] line ([OIII]/H$\beta <$ 3). Seyfert 1.9 galaxies have an emission line properties such that 
H$\alpha$ line is the only broad line component in their spectra \citep{osterbrock81}. The combined QSO and 
Type I sample, both featured in broad line components, makes up 55\% of our AGN sample. The fractions 
of Seyfert 1.9, Type II, and NLS1 in the sample are 26\%, 10\%, and 9\%, respectively. 
In Figure~\ref{fig21}, we provide some examples of representative optical images and spectra for each type.

The characteristic damping timescales of most AGNs (91\%) are longer than log $\tau =$ -1 while half of 
star-forming galaxies have log $\tau <$ -1. Therefore, optical variability will be a very efficient tool for 
separating AGNs from star-forming galaxies without emission line ratios diagnostics. We find some emission 
line galaxies showing spectral properties that are inconsistent with those of both star-forming galaxies and 
AGNs. Although the origin of their emission lines is ambiguous, it is unlikely that those emissions come from 
AGN activity based on their extremely short damping timescales (log $\tau <$ -1). Likewise, in composite 
galaxies where star formation coexists with AGN activity suggesting a connection between star formation and 
AGN \citep[e.g.,][]{heckman97, kewley06, davies07}, optical variability enables us to pinpoint AGN activity 
from composite galaxies that are ambiguous in the Baldwin-Phillips-Terlevich (BPT) diagnostics 
\citep{baldwin81, kewley06}. With the aid of optical variability information, we suggest that all our composite 
galaxies are AGN activity dominated. 

In the sample of 89 AGN candidates, there are contaminants by apparently normal (i.e., no emission lines)  
galaxies. Galaxies that are luminous in the X-ray but exhibit normal optical properties could be either heavily 
obscured AGNs (e.g., XBONGs candidates), accreting in a radiatively inefficient way, or true normal galaxies 
in X-ray emitting clusters or groups. In any cases, X-ray detection, as an additional constraint, is required to 
compile a complete sample of AGN candidates.   

We find 6 Type II AGN candidates. They are characterized by narrow emission lines without broad 
components, high-ionization line ratios, and hard X-ray spectra. No overlaps are found between our Type II
candidates and a list of Type II AGN candidates from \citet{zakamska03}. According to the unification model, 
none or very low level of variability is expected for Type II AGNs due to a highly obscured accretion disk from 
viewing angles \citep[e.g.,][]{yip09}. However, we detect non-negligible amount of optical 
variability in 2 of our Type II AGN candidates on long-term timescales consistent with typical or faint AGN. 
They could be Type II AGNs without severe attenuation either due to less dusty torus or marginal viewing 
angles. Alternatively, they could be `naked' Type I AGNs that lack broad line regions for some reason without 
signs of significant intrinsic absorption \citep{hawkins04}. 

For the first time, we conduct a systematic optical variability analysis for extended sources using a large 
difference image data set. Difference images allow us to reveal several different types of AGNs blended 
in extended sources, including Type II. However, it is impossible to take full advantage of such a large 
data set since there are observational (low S/N) and photometric (aperture photometry) 
limitations in this study.

\section{SUMMARY AND CONCLUSIONS}
\label{sec:concSec}
We quantify LCs of AGN candidates in Stripe 82 and identify AGNs based on their optical 
$g$-band variability. Our AGN candidates are sub samples of spectroscopically confirmed quasars (SpecQSO), 
photometrically selected quasars using optical colors (PhotoQSO), variable point sources (VarSrc), and 
X-ray detected sources (XRaySrc). To construct a photometric LC for an individual source, we use the 
recalibrated photometric data set of \citet{ivezic07}. We also construct difference image LCs by 
conducting a new image differencing and performing forced aperture photometry on all difference 
images. Although our difference image LCs have lower S/N compared to their counterpart photometric LCs 
due to aperture photometry, they have, on average, 20 more reliable data points by virtue of forced 
photometry on all difference images. For each LC, we estimate the standard deviation ($\Sigma$) and 
reduced $\chi^2$ ($\chi^2_{\nu}$) and confirm that most AGNs show intrinsic optical variability. 
We also characterize each LC with a damping timescale ($\tau$) and a variability amplitude 
($\hat{\sigma}$ for short-term or SF$_{\infty}$ for long-term). These SF parameters are derived in two 
different ways:(1) by directly fitting the first-order SF for the DRW (Equation~\ref{drwSF}) to all empirical 
SF($\Delta\,t$) data points, and (2) by modeling individual LCs as a DRW. We show that there is a linear 
relationship between the best-fits from the both methods, with a systematic difference in $\tau$ values in that 
those estimated from the simple direct fitting are smaller. We explore the impact of S/N and the number 
of data points in a LC on the SF parameter estimation. To provide a principled method for dealing with outliers 
in a LC and gauge how well the LC is described as the DRW, we perform jackknife resampling for each LC. 
For X-ray detected sources, we explore the relationship between the optical variability and the X-ray properties 
and also quantify the optical variability of extended sources in large survey for the first time.
Principal findings in our study are as follows:
\begin{enumerate}
\item We are able to select AGNs based on the optical variability detected in difference images with the high 
completeness and efficiency. For bright sources ($i <$ 19), we achieve C $=$ 98.6\%, 96.4\%, and 
82.2\% for log $\tau \ge$ 1.0, 1.5, and 2.0 criteria, respectively. With the sample including fainter sources 
($g <$ 20.5) as well, we achieve C $=$ 93.4\% and E $=$ 71.3\% for selection cuts on log $\tau \ge$ 
1.0 and log $\hat{\sigma} \le$ 4.15 by maximizing $\sqrt{C^2+E^2}$ .
\item We find some blazar candidates that are highly variable, with shorter $\tau$ than typical AGNs. One third 
of highly variable sources with IR colors consistent with known blazars are also radio detected.    
\item Since optical counterparts of the majority of X-ray detected sources are faint ($g >$ 21), their 
difference image LCs are mostly dominated by photometric noise that may spuriously induce apparent short 
damping timescales, mimicking those of non-variable objects. According to spectral classification, 
however, contamination rate by non-AGNs rapidly drops at $\tau >$ 0.1 days (less than 5\%). Thus, we 
are able to efficiently select faint AGNs from contaminants with a combination of additional X-ray information 
and a loose cut on $\tau$. 
\item No significant relationships are found between the rest-frame SF parameters and X-ray properties, 
such as hardness ratio and X-ray luminosity.      
\item Among X-ray detected optically--extended sources, we identify various types of AGNs 
including Type I, Type II, Seyfert 1.9, and NLS1 based on visual inspection of their optical spectra. 
\item For composite galaxies, we expect that the optical variability characteristics provides a powerful tool 
to determine whether a dominant emission source is AGN activity or star-forming.    
\item Contrary to the expectation from the unified model, one-third of our Type II AGNs show 
non-negligible variability on long-term timescales that are consistent with typical Type I AGNs.
\end{enumerate}

This study is an initial investigation of the effectiveness of difference image-based AGN classification. 
Future work can extend our analysis in several ways: (1) generating difference images with PSF-matched 
templates, (2) conducting PSF photometry on difference images, and (3) exploring all extended sources in 
Stripe 82. This will eventually lead to significant improvements in our understanding of properties of faint AGNs.

\acknowledgements
The authors thank Christopher S. Kochanek for helpful discussion. 
Support for this work was provided by the National Aeronautics and Space Administration through Chandra 
Award Numbers AR9-0015X, AR0-11014X, and AR2-13007X, issued by the Chandra X-ray Observatory 
Center, which is operated by the Smithsonian Astrophysical Observatory for and on behalf of the National 
Aeronautics Space Administration under contract NAS8-03060.

Funding for the SDSS and SDSS-II has been provided by the Alfred P. Sloan Foundation, the Participating Institutions, the National Science Foundation, the U.S. Department of Energy, the National Aeronautics and Space Administration, the Japanese Monbukagakusho, the Max Planck Society, and the Higher Education Funding Council for England. The SDSS Web Site is http://www.sdss.org/.

The SDSS is managed by the Astrophysical Research Consortium for the Participating Institutions. The Participating Institutions are the American Museum of Natural History, Astrophysical Institute Potsdam, University of Basel, University of Cambridge, Case Western Reserve University, University of Chicago, Drexel University, Fermilab, the Institute for Advanced Study, the Japan Participation Group, Johns Hopkins University, the Joint Institute for Nuclear Astrophysics, the Kavli Institute for Particle Astrophysics and Cosmology, the Korean Scientist Group, the Chinese Academy of Sciences (LAMOST), Los Alamos National Laboratory, the Max-Planck-Institute for Astronomy (MPIA), the Max-Planck-Institute for Astrophysics (MPA), New Mexico State University, Ohio State University, University of Pittsburgh, University of Portsmouth, Princeton University, the United States Naval Observatory, and the University of Washington.

This research made use of data obtained from the Chandra Source Catalog, provided by the 
Chandra X-ray Center (CXC) as part of the Chandra Data Archive.

\appendix
\section{Catalogs}
\subsection{Spectroscopically Confirmed Quasar Catalog}
\label{sec:SpecQsoCatSec}
\citet{schneider10} found 105,783 spectroscopically confirmed quasars from the SDSS DR7. They were 
selected  from quasar candidates (mainly color-selected) based on their spectral properties and a luminosity 
criterion ($M_{i} < -22.0$). The redshift distribution of the sample covers a wide range from 0.065 to 5.46, 
and has a median of $z\approx1.5$ \citep[see Figure 5 in][]{schneider10}. 
A total of 9,519 spectroscopically confirmed quasars (``{\bf SpecQSO}'') are located in 
Stripe 82 and approximately 97.2\% of them have matched photometric LCs from our SDSS 
photometry data set, recalibrated by \citet{ivezic07} for unresolved sources in Stripe 82 
(Section~\ref{sec:sDSSPhotoSec}). Even though this sample does not include very faint objects 
(having a mean $g$ magnitude of 19.9), it is very useful to study AGN variability because 
it is a {\it bona fide} quasar sample. Since the SpecQSO sample are bright and {\it bona fide}, 
many of these ($\sim$70\%) were also identified as optically variable sources by \citet{sesar07}
(Section~\ref{sec:VarCatSec}). In the $u$-$g$ vs. $g$-$r$ diagram (top left panel in Figure~\ref{fig2}), 
most SpecQSO occupy the bluer color portion (Region I) and some with high-redshift are most likely 
to occupy Region III.

\subsection{Photometrically Selected Quasar Catalog}
\label{sec:PhotoQsoCatSec}
In contrast to the SpecQSO catalog, the catalog of $\sim$1,000,000 SDSS DR6 quasar candidates, 
photometrically selected based on their colors \citep{richards09}, contains fainter sources as well. 
The mean $g$ magnitude of this sample is about 1 mag fainter than the SpecQSO sample. Thus it 
can be used to examine the capability of the difference imaging method at the faint end. Even at the 
bright end, it has more sources ($\sim$1.6 times more at $g <$ 18) than the SpecQSO sample since 
there are no fiber collision issues. The expected completeness of this photometric quasar catalog to type I 
quasars is about 70\% \citep{richards09}. The overall efficiency of their selection algorithm, which 
is much simpler compared to spectroscopic selection, reaches up to $\sim$70\%, while the 
efficiency for candidates flagged as ``good'' (i.e. a parameter ``good'' $\ge$ 0, ranging from -6 to 6) 
exceeds 90\%, especially at redshifts (z $\leqslant$ 2.2) where color contamination is 
minimal. Of the $\sim$1,000,000 objects in the catalog, 36,775 ``good'' quasar candidates 
( ``{\bf PhotoQSO}'') reside in Stripe 82 and also have photometric LCs. This is about 4 times larger 
than the SpecQSO sample, allowing for more quasar candidates even in Region III (top 
right panel in Figure~\ref{fig2}). It is worth noting that the contours for the PhotoQSO sample in 
Region I are a bit elongated toward redder colors, which is not seen for the SpecQSO sample. This 
might result from contamination by other sources, and we investigate possible causes of this 
elongation in Section~\ref{sec:Analysis}.

\subsection{Variable Source Catalog}
\label{sec:VarCatSec}
We use the Stripe 82 variable source catalog \citep{ivezic07, sesar07}.\footnote{Available at 
http://www.astro.washington.edu/users/ivezic/sdss/catalogs/S82variables.html} This catalog 
lists 67,507 variable point source candidates. They are selected by the following conditions:
\begin{itemize}
\item the number of observations in both the $g$ and $r$ bands $\ge$ 10
\item the root-mean-square (rms) scatter in both the $g$ and $r$ bands is $\ge$ 0.05 mag
\item the LC $\chi^{2}_{\nu}$ in both the $g$ and $r$ bands $\ge$ 3 around the mean
\item the median $g$ band mag is $<$ 20.5.
\end{itemize}
This catalog is dominated by the stellar locus stars (Region IV in Figure~\ref{fig2}) and low-redshift 
quasars (Region I). 
Based on colors, $\sim$15\% are expected to be low-redshift quasars (z $<$ 2) with some 
contamination from white dwarfs and binaries. The remaining sources are primarily main-sequence 
stars ($\sim$82\%) with some additional contribution from RR Lyrae (Region II; $\sim$1.75\%) 
and high-redshift quasars (Region III; $\sim$1.15\%). Based on the coverage of the SDSS Stripe 82
template images (from which we generate difference images), we select 31,992 sources from the original 
list of 67,507 variable sources. This sample is referred to as ``{\bf VarSrc}'', and all have photometric LCs. 
Since this VarSrc sample includes other types of variables than quasars, we are able study differences in the 
LCs' properties or SF parameters between quasars and other variable sources, such as stellar locus stars and 
RR Lyrae (Section~\ref{sec:Analysis}). 

\subsection{X-ray Source Catalog}
\label{sec:XrayCatSec}
The Chandra Source Catalog \citep[CSC;][]{evans10} provides X-ray properties for ~94,700 
distinct point and compact sources with observed spatial extent $\le$ 30$\arcsec$. 
Although this catalog only contains point-like X-ray sources, their optical counterparts may be 
extended sources such as host-dominated type I AGNs and type II AGNs. By removing optical point sources 
from the catalog, we compile a subset of AGN candidates that are bright and compact in the X-ray, but 
possibly extended in the optical. For the first time, we explore the optical variability of these X-ray 
luminous and optically resolved sources in Section~\ref{sec:resolvedSrcSec}. 
\citet{evans10} uniformly processed all the mission data from pointed observations and derived useful 
quantities, including aperture photometry fluxes in five energy bands, hardness ratio, and spectral model fits. 
To minimize the number of spurious sources in the catalog, they accepted only 3-sigma detections and set 
the minimum flux significance threshold to $\sim$10 photon counts in the broad band. This should lead to 
less than 1 false detection per image. There are about 1,900 X-ray sources within Stripe 82 
(``{\bf XRaySrc}''). 

To match the CSC with our photometric data set, we adopt a matching radius of 1$\arcsec$, 
corresponding to about $>$ 95\% match probability based on the statistical analysis of CSC-SDSS 
DR7 cross-match pairs \citep{rots11}. This yields only $\sim$640 matches, mostly because 
their optical counterparts are extended sources rather than unresolved sources: recall that our Stripe 
82 photometric data set does not include resolved sources (though in Section~\ref{sec:resolvedSrcSec}, 
we consider additional optically-resolved, X-ray compact sources in difference images). In addition, some of 
their optical counterparts are too faint to be detected by SDSS since X-ray selection generally favors objects
with higher redshift. The mean $g$ magnitude for matched sources are 21.33 (see Figure~\ref{fig1}). 
This sample is fainter than any other catalogs used in this study (down to $g \simeq$ 25).  

Although less than half of the XRaySrc have photometric LCs, we are still able to characterize 
the optical variability of all XRaySrc using difference image LCs. If we can address well how the optical 
variability of X-ray detected objects is related to their X-ray properties, it may be possible to identify 
AGNs in X-rays with high confidence even if they are not detected in optical 
(see Section~\ref{sec:identifyAGNByXRaysSec}).

\section{Fitting in Flux versus Magnitude}
%-------------- Fig 22 -----------------
\begin{figure} [tbp]
 \begin{center}
      \includegraphics[height=10cm]{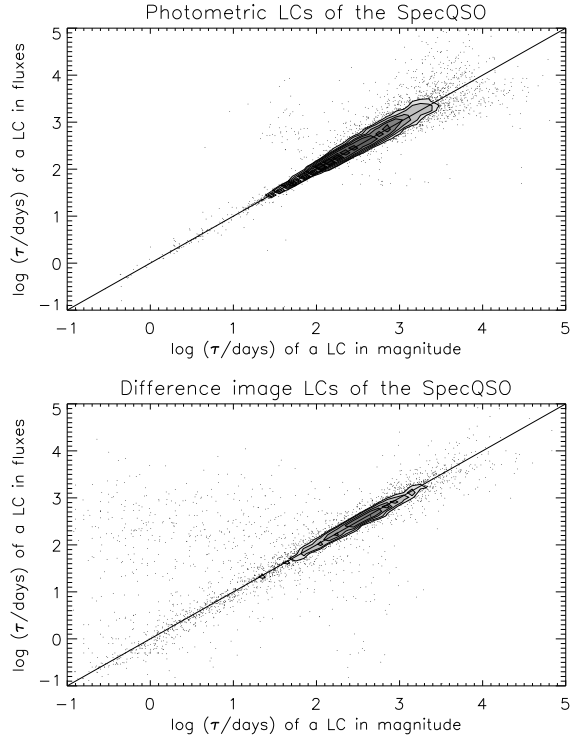}
      \caption{\label{appfig} Top panel: A self-consistency test of the direct fitting method against 
      the LC scales (linear vs. log) with photometric LCs. We overplot the one to one line (not a fit) to guide 
      the eye. 
      The result shows that the method is independent of the LC scales. 
      Bottom panel: Same, but for the DRW LC modeling with difference image LCs.}
   \end{center}
\end{figure}

To be able to compare difference image results with the previous analyses, which were conducted in magnitude, 
we test if the same characteristic timescales are obtained for a given LC analyzed both in magnitudes and in 
fluxes by a given LC fitting methodology. To perform this test, we convert photometric LCs of the SpecQSO 
from magnitude to flux, and then estimate the SF parameters based on flux (as well as magnitude) using the
direct fitting method. In Figure~\ref{appfig}, we compare log $\tau$ values for LCs in magnitudes to those for 
LCs in fluxes. For the direct fitting method, the slope is consistent with unity (0.993$\pm$0.014) and the 
y-intercept with zero (0.016$\pm$0.036). This good correlation indicates that the method is self-consistent 
against the linear and log scales at the observed low variability amplitudes. We also perform the same test
on the DRW LC modeling with difference image LCs of the SpecQSO. In previous studies of quasar photometric 
variability, the DRW model has been applied only to photometric LCs in magnitude \citep[e.g.,][]{kelly09, kozlowski10, macleod10, butler11}. In the conversion of difference image LCs from flux to magnitude, we 
add the template flux of a source to its difference image LC data points and take -2.5\,log of the total flux. 
However, we note that this conversion becomes unreliable in some cases where the absolute values of 
negative residual fluxes are comparable to or even greater than the template flux. This could happen if 
sources become intrinsically very faint. We exclude any LCs, which have at least one negative total flux, 
from this comparison. The test results using difference image LCs (Figure~\ref{appfig}) show that the DRW 
LC modeling is less self-consistent (a slope of 0.922$\pm$0.019 and an intercept of 0.323$\pm$0.042) 
compared to the direct fitting method. It has large scatter in the log $\tau <$ 1 regime. This scatter is mostly 
responsible for the positive y-intercept. In fact, one can expect an even tighter correlation than the direct fitting 
case by excluding these scattered points. However, we find no good reasons to exclude them, as they are 
sources with well-constrained LCs (see the following subsection). Thus we conclude that the direct fitting 
method is insensitive to the linear versus log scales for observed LCs of QSOs/AGNs, and shows a better 
self-consistency than the DRW LC modeling, especially in terms of the intercept.

% -------------- BEGIN BIBLIOGRAPHY -----------------

\bibliographystyle{apj3}
\bibliography{apj-jour,bibliography}

\clearpage

% -------------- BEGIN TABLES -----------------
\renewcommand{\arraystretch}{1.5}
%--------- Table 1 --------------------
\begin{table}
\begin{center}
\caption{List of catalogs for candidate AGNs used in this study} 
\label{catTable}
\begin{tabular}{c|c|c|c|c} \hline \hline
Object Type & Reference & Num. of sources     & Num. of sources  & Num. of sources\\
                 &                 & in Stripe 82 & with photometric LC  & with diffIm LC\\
\hline
SpecQSO  & \citet{schneider10} &  9,519  & 9,254  & 6,583\\
PhotoQSO & \citet{richards09}    & 36,884 & 36,775 & 24,596\\
VarSrc      & \citet{ivezic07}      & 31,992 & 31,992  & 21,196\\
XRaySrc   & \citet{evans10}     &  1,880  & 640      & 1,048\\
\hline
\end{tabular}
\tablecomments{The third column lists the total number of sources in Stripe 82. The fourth column lists the number of sources that are cross-matched with our Stripe 82 photometric light curve (LC) data set (point sources only). The last column shows the number of sources that have difference image LCs as well. In each catalog, the fraction of sources having difference image LCs is about two-thirds. This corresponds to the area fraction covered by camera columns 2-5 in Stripe 82.}
\end{center}
\end{table}

%--------- Table 2 --------------------
\begin{table}
\begin{center}
\caption{Overlap among catalogs}
\label{dupTable}
\begin{tabular}{c|c|c|c|c} \hline \hline
\diaghead{\theadfont Comparison Reference Catalog   }{Comparison Catalog}{Reference Catalog} & SpecQSO & PhotoQSO  & VarSrc  & XRaySrc\\
\hline
SpecQSO  & 1.000   &  0.221   & 0.097  & 0.078\\
PhotoQSO & 0.856   & 1.000    & 0.116  & 0.115\\
VarSrc      & 0.691   & 0.212   & 1.000   & 0.066\\
XRaySrc   & 0.015   & 0.006    & 0.002  & 1.000\\
\hline
\end{tabular}
\tablecomments{Fractions of sources in one reference catalog that are also found in the comparison catalog.}
\end{center}
\end{table}

%--------- Table 3 --------------------
\begin{table}
\begin{center}
\caption{Completeness for various $\tau$ criteria estimated for the SpecQSO sample}
\label{specQSOcomp}
\begin{tabular}{c|c|c|c} \hline \hline
Selection criterion & log ($\tau$/days) $\ge$ 1.0 & log ($\tau$/days) $\ge$ 1.5  & log ($\tau$/days) $\ge$ 2.0 \\
\hline
Bright sample ($i <$ 19) & 98.6\%    &  96.4\%                   & 82.2\%\\
Entire sample                 & 91.4\%    &  81.2\%                   & 56.7\%\\
\hline
\end{tabular}
\end{center}
\end{table}

%--------- Table 4 --------------------
\begin{table}
\begin{center}
\caption{Completeness and efficiency for various AGN selection criteria estimated for the VarSrc sample}
\label{varcomp}
\begin{tabular}{c|c|c} \hline \hline
Selection cuts & Completeness (\%) & Efficiency (\%) \\
\hline
log ($\tau$/days) $\ge$ 1.0, log ($\hat{\sigma}$/(counts/yr$^{1/2}$)) $\le$ 4.15 & 93.4 & 71.3 \\
log ($\tau$/days) $\ge$ 1.5, log ($\hat{\sigma}$/(counts/yr$^{1/2}$)) $\le$ 4.15 & 86.3 & 74.5 \\ 
log ($\tau$/days) $\ge$ 2.0, log ($\hat{\sigma}$/(counts/yr$^{1/2}$)) $\le$ 4.15 & 64.5 & 78.0 \\
\hline
\end{tabular}
\end{center}
\end{table}

\end{document}